\documentclass[aps,prb,twocolumn,showpacs,nobibnotes]{revtex4-1}

\usepackage{graphicx}
\usepackage{amsmath}
\usepackage{amssymb}
\usepackage{braket}
\usepackage{color}

\usepackage{hyperref} 
\hypersetup{
            colorlinks = true, 
            urlcolor   = blue, 
            linkcolor  = blue, 
            citecolor  = blue, 
            }

\begin{document}
\title{Efficient dielectric matrix calculations using the Lanczos algorithm for fast many-body $G_0W_0$ implementations}

\author{Jonathan Laflamme Janssen}
\affiliation{D\'epartement de physique et Regroupement Qu\'{e}b\'{e}cois sur les Mat\'{e}riaux de Pointe (RQMP), Universit\'e de Montr\'eal, C. P. 6128 Succursale Centre-ville, Montr\'eal (Qu\'ebec) H3C 3J7, Canada}
\author{Bruno~Rousseau}
\affiliation{D\'epartement de physique et Regroupement Qu\'{e}b\'{e}cois sur les Mat\'{e}riaux de Pointe (RQMP), Universit\'e de Montr\'eal, C. P. 6128 Succursale Centre-ville, Montr\'eal (Qu\'ebec) H3C 3J7, Canada}
\author{Michel C\^ot\'e}
\email[Corresponding author: ]{michel.cote@umontreal.ca}
\affiliation{D\'epartement de physique et Regroupement Qu\'{e}b\'{e}cois sur les Mat\'{e}riaux de Pointe (RQMP), Universit\'e de Montr\'eal, C. P. 6128 Succursale Centre-ville, Montr\'eal (Qu\'ebec) H3C 3J7, Canada}

\date{\today}

\begin{abstract}

We present a $G_0W_0$ implementation that assesses the two major bottlenecks of traditional plane-waves implementations, the summations over conduction states and the inversion of the dielectric matrix, without introducing new approximations in the formalism.
The first bottleneck is circumvented by converting the summations into Sternheimer equations.
Then, the novel avenue of expressing the dielectric matrix in a Lanczos basis is developed, which reduces the matrix size by orders of magnitude while being computationally efficient.
We also develop a model dielectric operator that allows us to further reduce the size of the dielectric matrix without accuracy loss. 
Furthermore, we develop a scheme that reduces the numerical cost of the contour deformation technique to the level of the lightest plasmon pole model. 
Finally, the use of the simplified quasi-minimal residual scheme in replacement of the conjugate gradients algorithm allows a direct evaluation of the $G_0W_0$ corrections at the desired real frequencies, without need for analytical continuation.
The performance of the resulting $G_0W_0$ implementation is demonstrated by comparison with a traditional plane-waves implementation, which reveals a 500-fold speedup for the silane molecule.
Finally, the accuracy of our $G_0W_0$ implementation is demonstrated by comparison with other $G_0W_0$ calculations and experimental results.

\end{abstract}


\maketitle

\section{Introduction}

Density functional theory (DFT)~\cite{Kohn:1965js,Capelle:2006tr,Martin} is currently the most popular approach for electronic structure simulations of periodic materials, molecules, and nanostructures. 
However, its predictive power is formally limited to ground-state properties. 
Consequently, while in practice DFT is widely used to calculate band structures, the precision of the results is limited. 
A formally sound and more precise~\cite{Onida02,Aulbur00} method is provided by the $GW$ framework~\cite{Hedin65,Hedin:1969wi}.
However, these calculations are computationally more demanding than their DFT counterparts.
Indeed, $GW$ calculations are typically limited to a few tens of atoms while DFT codes can handle a few hundreds. 
In a conventional plane-wave implementation~\cite{Hybertsen86}, two bottlenecks account for this limitation: the sums to be carried out over all conduction states~\cite{Stankovski:2011ej,Shih:2010ft,Friedrich:2011jr,Tamblyn:2011jx} and the inversion of the dielectric matrix~\cite{Umari10,Giustino10,Pham:2013gs} that describes the screening of an external potential by the simulated system.
Many different approaches have been explored to assess the summation over conduction states: it can be converted into a linear equation problem~\cite{b-24,Umari10,Lambert:2013bz,Pham:2013gs}, the conduction states can be replaced by simple approximate orbitals~\cite{Samsonidze:2011bu}, a so-called  extrapolar trick can be used to reduce the number of conduction states required for a given level of convergence~\cite{Bruneval:2008bw,Deslippe:2013fb}, the summations can be eliminated using the effective-energy technique~\cite{Berger:2010em,Berger:2012cx} or the size of the basis in which the Hamiltonian is expressed can be reduced through the use of localized basis sets~\cite{Blase11}.
Substantial attention has also been devoted to the assessment of the inversion of the dielectric matrix: it can be avoided by reformulating the problem into a self-consistent Sternheimer equation~\cite{Giustino10,Lambert:2013bz} or the size of the basis in which the dielectric matrix is expressed can be reduced either using Wannier orbitals~\cite{Umari09} or eigenvectors of the static dielectric matrix~\cite{Pham:2013gs}.

In this paper, we present a plane-wave implementation of the $G_0W_0$ method within the ABINIT project~\cite{Gonze:2009aa,Gonze:2005p10169,Gonze,Bottin:2008el,Goedecker:2006jn} that circumvents both bottlenecks.
The choice of plane-waves is motivated by its suitability for extended systems as well as its systematic convergence controlled by a single parameter.
In this implementation, to assess the summations over conduction states, we adopt the strategy to convert them into linear equation problems, since it is suitable for our choice of basis, efficient and well established~\cite{b-24,Umari10,Lambert:2013bz,Pham:2013gs}. 
To assess the bottleneck of the inversion of the dielectric matrix, we elaborate an approach where the matrix is expressed in a Lanczos basis~\cite{Golub}.
This reduces the size of the matrix as effectively as the traditional spectral decomposition method~\cite{Pham:2013gs,Wilson:2009p5717,Wilson:2008p6437} while being computationally an order of magnitude more efficient.
We also develop a model dielectric operator that allows to further reduce the size of the dielectric matrix. 

Furthermore, in the present $G_0W_0$ implementation, the use of the contour deformation technique~\cite{Giantomassi:2011ev,Lundqvist68} was preferred over plasmon pole models~\cite{Giantomassi:2011ev, Hybertsen86, Godby89, vonderLinden:1988ca, Engel:1993ia} to avoid considerations on the range of systems that can be accurately described~\cite{Stankovski:2011ej,Shaltaf:2008p7437}.
Traditionally, this choice implies a greater computational cost. 
In the present method, we explore two different directions to reduce this cost to the level of the simplest plasmon pole model~\cite{Hybertsen86}.
First, we use a Lorentzian to model the frequency dependence of the dielectric matrix and only treat the difference between this model and the exact dielectric matrix with the contour deformation technique, which alleviates the computational work required by the numerical integration.
This idea is inspired by previous work involving a Gaussian model~\cite{Anisimov:2000ub}, with the distinction of being compatible with the conversion of the summations over conduction states into linear equation problems and allowing a direct theoretical analogy with the plasmon pole technique.
Also, we elaborate a scheme to recycle the information computed in the construction of the static dielectric matrix and obtain the dynamical dielectric matrix at all relevant nonzero frequencies at a small computational cost, independent of the number of frequencies, thus reducing the numerical cost of the contour deformation technique to a level close to the simplest plasmon pole model~\cite{Hybertsen86}. 

The evaluation of the $G_0W_0$ corrections at the desired real frequency is usually unstable, due to the presence of poles on the real axis in the inverse dielectric matrix.
This difficulty is traditionally avoided by analytic continuation of the self-energy from the imaginary axis to the real axis~\cite{Rieger:1999p7438,Pham:2013gs,Lambert:2013bz,Umari10}.
In the present implementation, we solve this problem using a recently developed numerical method, the simplified quasi-minimal residual (SQMR) algorithm~\cite{Freund:1995vu}.
It is as efficient as the traditional conjugate gradients method, but stable when indefinite or nearly singular linear equations are solved, such as those involved in the calculation of the dielectric matrix at real frequencies. 

This article is organized as follows.
First, a summary of the $G_0W_0$ method is given in Sec.~\ref{reminder}.
Then, the bottleneck of the sums over conduction states is assessed in Sec.~\ref{sum_c}.
In particular, the Lorentzian model is developed in Sec.~\ref{self-energy}.
Next, the bottleneck of the dielectric matrix inversion is assessed in Sec.~\ref{Lanczos}.
Then, the model dielectric operator is developed in Sec.~\ref{model}.
A strategy to use the information generated in the construction of the static dielectric matrix to accelerate its computation at imaginary frequencies is devised in Sec.~\ref{recyclage}.
A theoretical analysis of the numerical cost of the present $G_0W_0$ implementation is given in Sec.~\ref{scaling}.
We compare our implementation with existing $G_0W_0$ schemes that also convert the summations over conduction states into Sternheimer equations in Sec.~\ref{relation}.
Then, we assess the accuracy of our implementation by comparing our results with previously published ones in Sec.~\ref{results}.
Finally, we assess its performance with respect to traditional implementations in Sec.~\ref{performance}.
Atomic units are used throughout unless otherwise specified.

\section{The $G_0W_0$ method}
\label{reminder}

In conventional implementations of the $G_0W_0$ method, corrections to DFT eigenenergies are obtained using first-order perturbation theory, where the perturbation is the difference between the $G_0W_0$ self-energy $\hat \Sigma(\omega)$ and the DFT exchange-correlation potential $\hat V^{\rm xc}$: 
\begin{equation}
\label{exc}
\Delta \varepsilon_e = \bra{e} \hat \Sigma(\varepsilon_e + \Delta \varepsilon_e) - \hat V^{\rm xc} \ket{e},
\end{equation}
where $\varepsilon_e$ is a DFT eigenenergy, $\ket{e}$ is the associated eigenstate, and $\Delta \varepsilon_e$ is the $G_0W_0$ correction to $\varepsilon_e$. 
The self-consistency with respect to $\Delta \varepsilon_e$ is easily avoided by making a Taylor expansion of Eq.~(\ref{exc}) to first order with respect to $\Delta \varepsilon_e$ around zero~\footnote
{
Once the Taylor expansion of Eq.~(\ref{exc}) is done, one can solve for $\Delta \varepsilon_e$.
The evaluation of $\bra{e} \hat \Sigma(\omega) \ket{e}$ at 2 frequencies then allows $\Delta \varepsilon_e$ to be obtained.
In the present implementation, we evaluate $\bra{e} \hat \Sigma(\omega) \ket{e}$ at 3 different energies to control the accuracy of the linear approximation for Eq.~(\ref{exc}).
We choose for these energies the values $\varepsilon_e$ (the DFT eigenvalue to be corrected),  $\varepsilon_e + \bra{e} \hat v_C - \hat V^{\rm xc} \ket{e}$ (the associated Hartree-Fock eigenvalue, evaluated using the bare Coulomb potential $\hat v_C$ and the DFT orbital) and their average.
This choice usually covers a range of energies nearly enclosing the corrected $G_0W_0$ eigenvalue and thus effectively controls the accuracy of the linearization of Eq.~(\ref{exc}).
}.
Also, the DFT exchange-correlation energy $\bra{e} \hat V^{\rm xc} \ket{e}$ can easily be extracted from the DFT calculation.
The only nontrivial part of the calculation is therefore the evaluation of the $G_0W_0$ exchange-correlation energy $\bra{e} \hat \Sigma(\omega) \ket{e}$, where $\hat \Sigma(\omega)$ is defined as
\begin{equation}
\begin{split}
\bra{r} & \hat \Sigma (\omega) \ket{r'} \\
& \equiv \frac{i}{2 \pi} \int_{-\infty}^{\infty} d \omega' e^{i \eta \omega'} \bra{r} \hat G_0(\omega' + \omega) \ket{r'} \bra{r} \hat W_0(\omega') \ket{r'},\\
\end{split}
\end{equation}
where $\hat G_0(\omega)$ is the Green's function, $\hat W_0(\omega)$ is the screened Coulomb potential, $\ket{r}$ is an eigenfunction of the position operator, and $\omega$ is the angular frequency (eventually $\in \mathbb{C}$).

In the following, we will restrict ourselves to the non-spin-polarized and nonperiodic (molecular) case for simplicity.
The Green's function is easily expressed in terms of DFT eigenstates $\ket{n}$ and eigenvalues $\varepsilon_n$,
\begin{equation}
\label{green}
\hat G_0(\omega) = \sum_{n=1}^\infty \frac{\ket{n} \bra{n}}{ \omega - \varepsilon_n + i \eta \, {\rm sgn}(\varepsilon_n-\mu)},
\end{equation}
where $\eta \to 0^+$ is a positive infinitesimal and $\mu$ is the chemical potential. 
The screened Coulomb potential is related to the bare Coulomb potential $\hat v_C$ by the inverse dielectric matrix $\hat \epsilon^{-1}(\omega)$,
\begin{equation}
\hat W_0(\omega) = \hat v_C^{1/2} \hat \epsilon^{-1}(\omega) \hat v_C^{1/2},
\end{equation}
where $\hat v_C^{1/2}$ is the square root of the Coulomb potential.
We choose to work here with the symmetric form of the dielectric matrix because of its computational advantages~\footnote
{
The symmetric form of the dielectric matrix~\cite{Giantomassi:2011ev,Giustino10,Pham:2013gs} $\hat \epsilon(\omega) = 1 - \hat v_C^{1/2} \hat P(\omega) \hat v_C^{1/2}$, where $\hat P(\omega)$ is the irreducible polarizability, is Hermitian, in contrast to its usual form $\hat \epsilon(\omega) = 1 - \hat v_C \hat P(\omega)$.
This makes it possible to apply the matrix on a vector only one time per iteration instead of two in Lanczos based linear algebra methods~\cite{Freund:1995vu}.
Indeed, in the present $G_0W_0$ implementation, a substantial part of the computation time is spent on the application of the dielectric matrix within the Lanczos algorithm. 
Using this form of the matrix thus halves this part of the computation time, which provides a substantial numerical advantage.
}.
These definitions allow us to express the $G_0W_0$ exchange-correlation energy as 
\begin{equation}
\label{e_xc}
\begin{split}
& \Sigma_e (\delta) \equiv \bra{e} \hat \Sigma(\varepsilon_e+\delta) \ket{e} \\
& \equiv \frac{i}{2 \pi} \sum_{n=1}^\infty \int_{-\infty}^{\infty} d \omega e^{i \eta \omega} \frac{\bra{en^*} \hat v_C^{1/2} \hat \epsilon^{-1}(\omega) \hat v_C^{1/2} \ket{n^*e}}{\omega - \omega_{ne} + i \eta \, {\rm sgn}(\varepsilon_n - \mu)},
\end{split}
\end{equation}
where $\omega_{ne}\equiv\varepsilon_n-\varepsilon_e-\delta$, $\braket{r|n^*} \equiv \braket{r|n}^*$ and $\braket{r | n^*e} \equiv \braket{r|n}^* \braket{r|e}$. 
It is customary at this stage to split the self-energy matrix element into an exchange $\Sigma^{\rm x}_e (\delta)$ and a correlation $\Sigma^{\rm c}_e (\delta)$ part,
\begin{equation}
\label{e_x1}
\begin{split}
& \Sigma^{\rm x}_e (\delta) \equiv \bra{e} \hat \Sigma^{\rm x}(\varepsilon_e+\delta) \ket{e} \\
& \equiv \frac{i}{2 \pi} \sum_{n=1}^\infty \int_{-\infty}^{\infty} d \omega e^{i \eta \omega} \frac{\bra{en^*} \hat v_C^{1/2} \, \hat 1 \, \hat v_C^{1/2} \ket{n^*e}}{\omega - \omega_{ne} + i \eta \, {\rm sgn}(\varepsilon_n - \mu)}, 
\end{split}
\end{equation}
and
\begin{equation}
\label{e_c1}
\begin{split}
& \Sigma^{\rm c}_e (\delta) \equiv \bra{e} \hat \Sigma^{\rm c}(\varepsilon_e+\delta) \ket{e} = \\
&\frac{i}{2 \pi} \sum_{n=1}^\infty \int_{-\infty}^{\infty} d \omega e^{i \eta \omega} \frac{\bra{en^*} \hat v_C^{1/2} ( \hat \epsilon^{-1}(\omega) - \hat 1 ) \hat v_C^{1/2} \ket{n^*e}}{\omega - \omega_{ne} + i \eta \, {\rm sgn}(\varepsilon_n - \mu)}.
\end{split}
\end{equation}
The integral in the exchange part can be evaluated by closing the integration contour with a half-circle of infinite radius in the upper complex plane.
The factor $e^{i \eta \omega}$ reduces the integral over this half-circle to 0 and the residues of the poles included in the contour become the only contributions to the exchange term $\Sigma^{\rm x}_e$.
Since the presence of a pole above or below the real axis is determined by the presence of its eigenenergy below or above the chemical potential $\mu$ [see Eq.~(\ref{e_x1})], the poles included in the contour are those associated to the valence states. 
The exchange term thus takes the familiar form
 \begin{equation}
\label{e_x}
\Sigma^{\rm x}_e = \bra{e} \hat \Sigma^{\rm x} \ket{e} = - \sum_v \bra{ev^*} \hat v_C \ket{v^*e},
\end{equation}
where the index $v$ labels the valence states.

\begin{figure}
\includegraphics[width=1.0 \linewidth]{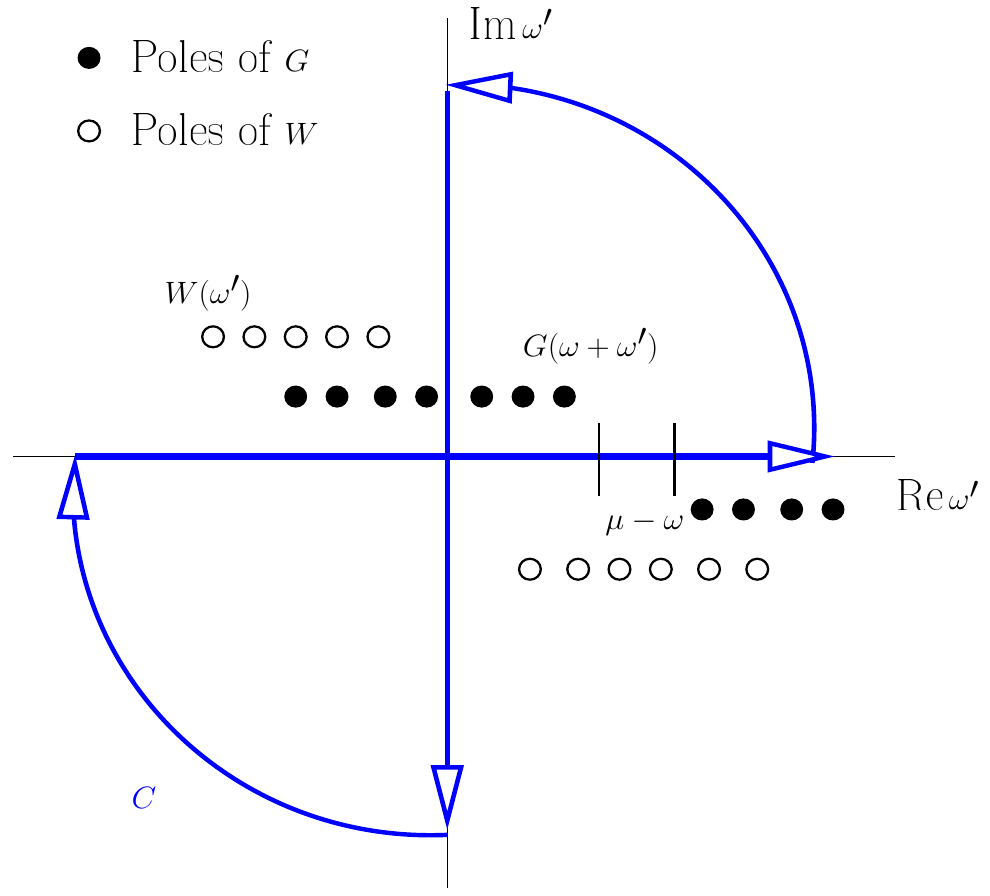} 
\caption[The path used in the contour deformation technique]{ \label{contour} The path used in the contour deformation technique. The poles of the screened Coulomb interaction $\hat W_0(\omega)$ lie outside the contour, only some poles of the Green's function $\hat G_0(\omega)$ lie inside. This figure is reproduced with permission from Ref.~\onlinecite{Giantomassi:2011ev}.}
\end{figure}

A popular way to calculate the integration over $\omega$ in $\Sigma^{\rm c}_e (\delta)$ is the plasmon-pole approximation~\cite{Giantomassi:2011ev, Hybertsen86, Godby89, vonderLinden:1988ca, Engel:1993ia}, which reduces the associated computational cost by only requiring the explicit calculation of the dielectric matrix at one or two frequencies. 
However, the range of systems where this approximation is robust is subject to some debate~\cite{Stankovski:2011ej,Shaltaf:2008p7437}. 
To preserve the precision and the wide applicability (in terms of physical systems) of the plane-wave basis set, we instead carry out the integration numerically. 
However, the high number of poles along the real axis makes it unwieldy to integrate numerically along this direction. 
Therefore we use the residue theorem to reformulate the problem into an integration along the imaginary axis, using the contour illustrated in Fig.~\ref{contour}~\cite{Giantomassi:2011ev,Anisimov:2000ub,Lundqvist68,Godby88}.

The correlation part then becomes
\begin{equation}
\label{e_c2}
\begin{split}
\Sigma^{\rm c}_e (\delta) = & \frac{-i}{2 \pi} \sum_{n=1}^\infty \Bigg( \int_{C1} + \int_{-i\infty}^{i\infty} + \int_{C3} \Bigg) \\
& \times d \omega \frac{\bra{en^*} \hat v_C^{1/2} \big( \hat \epsilon^{-1}(\omega) - \hat 1 \big) \hat v_C^{1/2} \ket{n^*e}}{\omega - \omega_{ne}} \\
& - \sum_v \bra{ev^*} \hat v_C^{1/2} \big( \hat \epsilon^{-1}(\omega_{ve}) - \hat 1 \big) \hat v_C^{1/2} \ket{v^*e} \Theta(\omega_{ve}) \\
& + \sum_c \bra{ec^*} \hat v_C^{1/2} \big( \hat \epsilon^{-1}(\omega_{ce}) - \hat 1 \big) \hat v_C^{1/2} \ket{c^*e} \Theta(-\omega_{ce}),
\end{split}
\end{equation}
where we took the limit $\eta \to 0^+$ after selecting the poles lying inside the contour and where the index $c$ labels the conduction states.
It can be shown that $\lim_{|\omega| \to \infty} \big( \hat \epsilon^{-1}(\omega) - \hat 1 \big) \to 0$ as $1/\omega^2$.
Consequently, the integrals over both quarters of circle ($\int_{C1}$ and $\int_{C3}$) vanish.
By substituting $\omega \to i\omega$, the domain of integration of the second term of Eq.~(\ref{e_c2}) can be made real,
\begin{equation}
\label{e_c3}
\begin{split}
\Sigma^{\rm c}_e (\delta) = & \frac{1}{\pi} \sum_{n=1}^\infty \int_{0}^{\infty} d \omega \bra{en^*} \hat v_C^{1/2} \big( \hat \epsilon^{-1}(i \omega) - \hat 1 \big) \hat v_C^{1/2} \ket{n^*e} \\
& \times \frac{\omega_{ne}}{\omega^2 + \omega_{ne}^2} + \Sigma^{\rm P}_e(\delta),
\end{split}
\end{equation}
where we have defined the residue term
\begin{equation}
\label{pe}
\begin{split}
\Sigma^{\rm P}_e (\delta) \equiv & - \sum_{v} \bra{ev^*} \hat v_C^{1/2} \big( \hat \epsilon^{-1}(\omega_{ve}) - \hat 1 \big) \hat v_C^{1/2} \ket{v^*e} \Theta(\omega_{ve}) \\
& + \sum_{c} \bra{ec^*} \hat v_C^{1/2} \big( \hat \epsilon^{-1}(\omega_{ce}) - \hat 1 \big) \hat v_C^{1/2} \ket{c^*e} \Theta(-\omega_{ce}).
\end{split}
\end{equation}

The matrix element $\bra{en^*} \hat v_C^{1/2} \big( \hat \epsilon^{-1}(i\omega) - \hat 1 \big) \hat v_C^{1/2} \ket{n^*e}$ in the integral over frequencies varies slowly as a function of $\omega$ with respect to the Lorentzian $\omega_{ne}/(\omega^2 + \omega_{ne}^2)$.
Therefore, the integral can be made smoother by subtracting a properly normalized Lorentzian from the integrand and carrying out its integration analytically.
This step turns out to be necessary when $\omega_{ne} \to 0$, e.g., when a pole of the Green's function lies on the imaginary axis, since the Lorentzian then becomes a Dirac delta and its numerical integration then becomes problematic. 
This idea can be refined by multiplying the Lorentzian by a scalar function $f(\omega)$ that models approximately the frequency dependence of the dielectric matrix, which makes the integrand smaller and thus easier to sample numerically~\cite{Anisimov:2000ub}.
The resulting expression is
\begin{equation}
\label{e_c}
\begin{split}
\Sigma^{\rm c}_e & (\delta) \\
= & \frac{1}{\pi} \int_0^\infty d \omega \Big( \sigma^{\rm N}_e(i\omega,\delta) - \sigma^{\rm N0}_e(i\omega,\delta) \Big) + \Sigma^{\rm A}_e(\delta) + \Sigma^{\rm P}_e(\delta),
\end{split}
\end{equation}
where
\begin{equation}
\label{dge}
\begin{split}
\sigma^{\rm N}_e(i\omega,\delta) \equiv \sum_n & \bra{en^*} \hat v_C^{1/2} \big( \hat \epsilon^{-1}(i \omega) - \hat 1 \big) \hat v_C^{1/2} \ket{n^*e} \\
& \times \frac{\omega_{ne}}{\omega^2 + \omega_{ne}^2} ,
\end{split}
\end{equation}
\begin{equation}
\label{dg0e}
\begin{split}
\sigma^{\rm N0}_e(i\omega,\delta) \equiv \sum_n & \bra{en^*} \hat v_C^{1/2} \big( \hat \epsilon^{-1}(0) - \hat 1 \big) \hat v_C^{1/2} \ket{n^*e} \\
& \times f(\omega) \frac{\omega_{ne}}{\omega^2 + \omega_{ne}^2},
\end{split}
\end{equation}
\begin{equation}
\label{dhe}
\begin{split}
\Sigma^{\rm A}_e(\delta) \equiv & \frac{1}{\pi} \int_0^\infty d\omega \, \sigma^{\rm N0}_e(i\omega,\delta) \\
= & \sum_{n} \bra{en^*} \hat v_C^{1/2} \big( \hat \epsilon^{-1}(0) - \hat 1 \big) \hat v_C^{1/2} \ket{n^*e} F_{ne}(\delta),
\end{split}
\end{equation}
and where
\begin{equation}
\label {Fne}
F_{ne}(\delta) \equiv \frac{1}{\pi} \int_0^\infty d\omega f(\omega) \frac{\omega_{ne}}{\omega^2 + \omega_{ne}^2}.
\end{equation}
In this form, the numerical integration required by Eq.~(\ref{e_c}) becomes straightforward~\footnote
{ 
A simple change of variable $\omega = (1+t)/(1-t)$ allows to remap the semi-infinite interval $[0,+\infty[$ to $]-1,1]$.
The integration can now be carried out numerically using methods such as Gauss-Legendre quadrature.
We found that a sampling of about $10$ frequencies converges the integral to 1~meV, in agreement with previous work~\cite{Blase11,Godby89}. 
}.

Equations~(\ref{pe})-(\ref{Fne}) contain quantities that can all be obtained from DFT calculations (eigenvalues $\varepsilon_n$ and eigenfunctions $\ket{n}$), except for the inverse dielectric matrix $\hat \epsilon^{-1}(\omega)$.
The latter is usually obtained by direct inversion of the dielectric matrix in some basis. 
The required dielectric matrix elements can be obtained from DFT eigenvalues and eigenfunctions using the Adler-Wiser expression for the random phase approximation to the irreducible polarizability~\cite{Hedin:1969wi,Adler:1962p6392,Wiser:1963p6405}
\begin{equation}
\label{epsilon}
\hat \epsilon(\omega) = 1 - \hat v_C^{1/2} \hat P(\omega) \hat v_C^{1/2},
\end{equation}
\begin{equation}
\label{P}
\begin{split}
\hat P(\omega) = 2 \sum_{cv} \ket{c^*v} \bigg[ & \frac{1}{\omega - (\varepsilon_c - \varepsilon_v)} - \frac{1}{\omega + (\varepsilon_c - \varepsilon_v)} \bigg] \bra{vc^*}.
\end{split}
\end{equation}

The preceding equations suffice to describe conventional $G_0W_0$ calculations in a plane-wave basis set, which involve calculating a sufficient number of conduction states to converge the summations in Eqs.~(\ref{dge}), (\ref{dg0e}), (\ref{dhe}), and~(\ref{P}) as well as inverting $\hat \epsilon(\omega)$ in a plane-waves basis.

\section{Avoiding summations over conduction states}
\label{sum_c}

The bottleneck of the summation over conduction states can be avoided without introducing further approximations to the preceding scheme at the expense of introducing a linear equation problem to be solved iteratively~\cite{b-24,Umari10,Lambert:2013bz,Pham:2013gs}.
This strategy is commonly used in density functional perturbation theory~\cite{Baroni01,Gonze:1995p6284}, where this type of linear equation is frequently referred to as a \emph{Sternheimer equation}~\cite{Sternheimer:1954hm}.
First, we will apply this idea to the polarizability.
Then, we will eliminate the summation over states present in Eqs.~(\ref{dge})-(\ref{dhe}) as well. 

\subsection{The polarizability}
We start from the action of the polarizability on some vector $\ket{\psi_j}$, labeled by the index $j$, 
\begin{equation}
\label{P1}
\begin{split}
\hat P (\omega) \ket{\psi_j} = & - 2 \sum_v \ket{v} \bigg( \sum_c \ket{c^*} \frac{1}{\varepsilon_c - \varepsilon_v - \omega} \braket{vc^* |\psi_j} \\
& + \sum_c \ket{c^*} \frac{1}{\varepsilon_c - \varepsilon_v + \omega} \braket{vc^* | \psi_j} \bigg), \\
= & - 2 \sum_v \ket{v} \Big( \ket{f^*_{jv-}(\omega)} + \ket{f^*_{jv+}(\omega)} \Big),
\end{split}
\end{equation}
where $\ket{v} \ket{f^*_{jv\pm}(\omega)} = \ket{v f^*_{jv\pm}(\omega)}$ and where we have introduced the new vector
\begin{equation}
\label{fvpm}
\ket{f_{jv\pm}(\omega)} = \sum_c \ket{c} \frac{1}{\varepsilon_c - \varepsilon_v \pm \omega} \braket{c | v \psi_j^*}.
\end{equation}
The idea is simply to use the completeness rule $\sum_c \ket{c} \bra{c} = 1 - \sum_v \ket{v} \bra{v} \equiv \hat {\mathcal P}_c$, which can readily be done if $\varepsilon_c$ is substituted by $\hat H$ in the denominator of Eq.~(\ref{fvpm}).
We thus obtain
\begin{equation}
\label{P2a}
\ket{f_{jv\pm}(\omega)} = \frac{\hat {\mathcal P}_c}{\hat H - \varepsilon_v \pm \omega} \ket{v\psi_j^*}.
\end{equation}
However, inverting the Hamiltonian is a problem similar in size to its full diagonalization, e.g. the calculation of all its eigenvalues and eigenvectors, which is exactly what we set out to avoid in the first place. 
A numerically less expensive alternative is to turn the problem into a linear equation,
\begin{equation}
\label{P2}
\Big( \hat H - \varepsilon_v \pm \omega \Big) \ket{f_{jv\pm}(\omega)} = \hat {\mathcal P}_c \ket{v\psi_j^*}.
\end{equation}
This equation becomes problematic to solve when $\omega \to 0$, since the left-hand side operator becomes singular.
The customary way to avoid this instability is to add a term $\beta {\mathcal P}_v$ to the Hamiltonian $\hat H$, where $\beta$ is larger than the valence bandwidth.
Thus, when $\omega \to 0$, the operator acting on $\ket{f_{jv\pm}(\omega)}$ does not become singular~\cite{Baroni01}. 
Since the right-hand side of Eq.~(\ref{P2}) is outside the valence subspace, this algebraic trick leaves the final answer unaffected.
However, when the argument of the dielectric operator $\omega$ is real, as it is the case in Eq.~(\ref{pe}), this trick does not prevent the left-hand side from being singular. 
Indeed, $\delta$ could be chosen so that $\omega_{ve}$, for some $\varepsilon_v > \varepsilon_e+\delta$ (or $\omega_{ce}$, for some $\varepsilon_c < \varepsilon_e+\delta$) equals some valence-conduction transition $\omega_{c'v'}$ (or $-\omega_{c'v'}$). 
Then, the operator $\hat H - \varepsilon_{v'} - \omega_{ve}$ (or $\hat H - \varepsilon_{v'} + \omega_{ce}$) would become singular in some part of the conduction subspace.
Since the right-hand side of Eq.~(\ref{P2}) can have nonzero components in this subspace, the previous trick cannot be applied to eliminate the singularity. 
Indeed, there will be an associated subspace where $\hat \epsilon(\omega_{ce})$ (or $\hat \epsilon(\omega_{ve})$) will be infinite and where $1 - \hat \epsilon^{-1}$ will be $1$, so that the kernel of $\hat H - \varepsilon_{v'} \pm \omega$ will contribute to $\Sigma^{\rm P}_e(\delta)$ as per Eq.~\eqref{pe}. 
However, it is still possible to stabilize Eq.~(\ref{P2}) without altering this physical contribution by using the \emph{Simplified Quasi-Minimal Residual} (SQMR) algorithm~\cite{Freund:1995vu} instead of conjugate gradients~\cite{Golub,Saad}.
Indeed, the former is stable for indefinite matrices close to singularity ($\bra{n} \hat H - \varepsilon_v \pm \omega \ket{n} \gtrsim 10^{-3}$ Ha $\forall \ket{n}$) while the latter is stable only for positive definite matrices~\footnote
{
Both SQMR and conjugate gradients algorithms solve approximately the linear equation:
\begin{equation*}
\hat A \ket{x} = \ket{b}, 
\end{equation*}
where $\hat A$ is an Hermitian operator, $\ket{b}$ is a vector and $\ket{x}$ is the solution to the linear equation.
More precisely, at the $N^{\rm th}$ iteration, they have constructed an approximate solution $\ket{x_N}$, which is the exact solution of the linear equation projected in the $N$-dimension Krylov subspace
\begin{equation*}
\{ \ket{b}, \hat A \ket{b}, \hat A^2 \ket{b}, ..., \hat A^{N-1} \ket{b} \}.
\end{equation*}
The key advantage of these methods is that they do not require the whole basis describing the Krylov subspace to be stored in memory. 
Instead, they use recursion relations where the solution $\ket{x_{N+1}}$ is expressed in term of $\ket{x_{N}}$ and a small constant number of other vectors. 
In the specific case of conjugate gradients, a $LDL^T$ factorization~\cite{Golub} is used to obtain such a recursion relation, which requires the operator $\hat A$ to be positive definite.
The SQMR method circumvent this requirement by replacing the $LDL^T$ factorization by a $QR$ factorization~\cite{Golub}, which only requires the operator $\hat A$ to be nonsingular but increases slightly the computational cost. 
Since the SQMR and the MINRES methods are mathematically equivalent when no preconditioning is used~\cite{Freund:1995vu}, the interested reader may consult the original work on the MINRES method~\cite{Paige:1975jv} for a more detailed explanation of the method. 
}.
It thus becomes easy to choose a suitable value of $\delta$ such that Eq.~(\ref{P2}) is stable, without adding substantially to the computation time.
Still, for the important case $\omega=0$, the operator will be singular in the valence subspace up to machine precision.
Therefore, the addition of $\beta {\mathcal P}_v$ to $\hat H - \varepsilon_v \pm \omega$ remains necessary. 
In our implementation, we adopted the equivalent strategy to orthogonalize the solution vector with respect to the valence subspace at each SQMR step.
This SQMR-based scheme has the advantage of allowing the direct calculation of $\Sigma^{\rm c}_e (\delta)$ at the (real) desired frequency, without requiring the addition of an imaginary infinitesimal to the Hamiltonian in Eq.~\eqref{P2} and the associated convergence study~\footnote{
See the \emph{zcut} input variable documentation on the ABINIT website~\cite{ABINIT:site}.}
nor the use of analytic continuation~\cite{Pham:2013gs} and related stability considerations~\cite{Rieger:1999p7438}.

We note that for imaginary frequencies $i\omega$, $\hat H - \varepsilon_v \pm i\omega$ is not Hermitian, which prevents the use of SQMR or conjugate gradients.
However, in this case, we can instead solve the Hermitian linear equation
\begin{equation}
\label{P2+}
\Big( (\hat H - \varepsilon_v)^2 - (i\omega)^2 \Big) \ket{f_{jv}(i\omega)} = 2(\hat H - \varepsilon_v) \hat {\mathcal P}_c \ket{v\psi_j^*},
\end{equation}
where the solution obtained is the sum of the vectors defined in Eq.~(\ref{P2a}),
\begin{equation}
\label{fv}
\ket{f_{jv}(\omega)} \equiv \ket{f_{jv+}(\omega)} + \ket{f_{jv-}(\omega)}.
\end{equation}
Typically, solving this equation to a precision where the residual $|\Big( (\hat H - \varepsilon_v)^2 - (i\omega)^2 \Big) \ket{f_{jv}(i\omega)} - 2(\hat H - \varepsilon_v) \hat {\mathcal P}_c \ket{v\psi_j^*}|$ is less than $10^{-10}$~Ha$^2$ converges the calculated $\Sigma^{\rm c}_e (\delta)$ to 5~meV.

\subsection{The self-energy}
\label{self-energy}
Similarly to the case of the polarizability, we use the completeness rule $\sum_n \ket{n} \bra{n} = 1$ to eliminate the summation over all states in $\Sigma^{\rm c}_e (\delta)$.
However, it is now necessary to introduce an intermediate basis that we will refer to as $\{ \ket{l} \}$. 
We only require that this basis be complete, up to a convergence criteria on $\Sigma^{\rm c}_e (\delta)$.
For $\sigma^{\rm N}_e(i\omega,\delta)$, we obtain from Eq.~(\ref{dge}),
\begin{equation}
\begin{split}
\sigma^{\rm N}_e(i\omega,\delta) = & \sum_{n,l,l'} \bra{en^*} \hat v_C^{1/2} \ket{l} \bra{l} \big( \hat \epsilon^{-1}(i \omega) - \hat 1 \big) \ket{l'} \\
& \times \bra{l'} \hat v_C^{1/2} \ket{n^*e} \frac{\omega_{ne}}{\omega^2 + \omega_{ne}^2}, \\
= & \sum_{n,l,l'} \bra{l} \big( \hat \epsilon^{-1}(i \omega) - \hat 1 \big) \ket{l'} \\
& \times \bra{l^*} \hat v_C^{1/2} \ket{e^*n} \frac{\omega_{ne}}{\omega^2 + \omega_{ne}^2} \bra{ne^*} \hat v_C^{1/2} \ket{l'^*}.
\end{split}
\end{equation}
Let $\hat \Phi_e$ be an operator so that
\begin{equation}
\bra{\phi} \hat \Phi_e \ket{r} \equiv \phi^*(r) \phi_e(r)
\end{equation}
Then, $\sigma^{\rm N}_e(i\omega,\delta)$ becomes
\begin{equation}
\begin{split}
\sigma^{\rm N}_e(i\omega,\delta) = & \sum_{n,l,l'} \bra{l} \big( \hat \epsilon^{-1}(i \omega) - \hat 1 \big) \ket{l'} \\
& \times \bra{l^*} \hat v_C^{1/2} \hat \Phi_e^\dagger \ket{n} \frac{\omega_{ne}}{\omega^2 + \omega_{ne}^2} \bra{n} \hat \Phi_e \hat v_C^{1/2} \ket{l'^*}.
\end{split}
\end{equation}
Similarly to the case of the polarizability, we need to replace the eigenvalues $\varepsilon_n$ by the Hamiltonian $\hat H$ to use the completeness relation,
\begin{equation}
\begin{split}
\sigma^{\rm N}_e(i\omega,\delta) = & \sum_{l,l'} \bra{l} \big( \hat \epsilon^{-1}(i \omega) - \hat 1 \big) \ket{l'} \bra{l^*} \hat v_C^{1/2} \hat \Phi_e^\dagger \\
& \times \frac{\hat H - \varepsilon_e-\delta}{\omega^2 + (\hat H - \varepsilon_e-\delta)^2} \hat \Phi_e \hat v_C^{1/2} \ket{l'^*}.
\end{split}
\end{equation}
To avoid inverting the Hamiltonian, we need to transform the problem into a linear equation. 
We can do this easily by defining the vector $\ket{\sigma^{\rm N}_{e,l'}(i\omega,\delta)}$,
\begin{equation}
\label{ge}
\begin{split}
\sigma^{\rm N}_e(i\omega,\delta) = & \sum_{l,l'} \bra{l} \big( \hat \epsilon^{-1}(i \omega) - \hat 1 \big) \ket{l'} \\
& \times \bra{l^*} \hat v_C^{1/2} \hat \Phi_e^\dagger \ket{\sigma^{\rm N}_{e,l'}(i\omega,\delta)},
\end{split}
\end{equation}
which is given by the following linear equation,
\begin{equation}
\label{kge}
\big( \omega^2 + (\hat H - \varepsilon_e-\delta)^2 \big) \ket{\sigma^{\rm N}_{e,l'}(i\omega,\delta)} = ( \hat H - \varepsilon_e-\delta ) \hat \Phi_e \hat v_C^{1/2} \ket{l'^*}.
\end{equation}
Similarly, for $\sigma^{\rm N0}_e(i\omega,\delta)$, we obtain
\begin{equation}
\label{g0e}
\begin{split}
\sigma^{\rm N0}_e(i\omega,\delta) = & \sum_{l,l'} \bra{l} \big( \hat \epsilon^{-1}(0) - \hat 1 \big) \ket{l'} f(\omega) \\
& \times \bra{l^*} \hat v_C^{1/2} \hat \Phi_e^\dagger \ket{\sigma^{\rm N}_{e,l'}(i\omega,\delta)}.
\end{split}
\end{equation}
The same strategy can also be applied to eliminate the summation over all states in $\Sigma^{\rm A}_e(\delta)$:
\begin{equation*}
\begin{split}
\Sigma^{\rm A}_e(\delta) = & \sum_{n,l,l'} \bra{l} \big( \hat \epsilon^{-1}(0) - \hat 1 \big) \ket{l'} \\
& \times \bra{l^*} \hat v_C^{1/2} \hat \Phi_e^\dagger \ket{n} F_{ne}(\delta) \bra{n} \hat \Phi_e \hat v_C^{1/2} \ket{l'^*}.
\end{split}
\end{equation*}
To go further, we now need to know the explicit expression for $F_{ne}(\delta)$ and, therefore, to choose the model function $f(\omega)$.
The obvious minimalist case is to not attempt any modelization of the frequency dependence of $\hat \epsilon^{-1}(i \omega) - \hat 1$.
We then have
\begin{equation}
\label{f1}
\begin{aligned}
f(\omega) =& 1, \\ 
F_{ne}(\delta) =& \frac{1}{2} \big( \Theta(\omega_{ne}) - \Theta(-\omega_{ne}) \big), \\
\Sigma^{\rm A}_e(\delta) =& \frac{1}{2} \sum_{l,l'} \bra{l} \big( \hat \epsilon^{-1}(0) - \hat 1 \big) \ket{l'} \\
& \times \bra{l^*} \hat v_C^{1/2} \hat \Phi_e^\dagger \big(\hat {\mathcal Q}_e(\delta) - \hat {\mathcal P}_e(\delta) \big) \hat \Phi_e \hat v_C^{1/2} \ket{l'^*}, \\
\end{aligned}
\end{equation}
where 
\begin{equation}
\label{QePe}
\begin{aligned}
\hat {\mathcal Q}_e(\delta) \equiv & \sum_{\varepsilon_n > \varepsilon_e+\delta} \ket{n} \bra{n}, \\
\hat {\mathcal P}_e(\delta) \equiv & \sum_{\varepsilon_n < \varepsilon_e+\delta} \ket{n} \bra{n}. \\
\end{aligned}
\end{equation}

We note the presence of discontinuities with respect to $\delta$ in $\Sigma^{\rm A}_e(\delta)$ and $\Sigma^{\rm P}_e (\delta)$ [see Eq.~(\ref{pe})], which must be treated carefully so that $\Sigma^{\rm c}_e(\delta)$ remains continuous~\footnote
{ 
Discontinuities in $\Sigma^{\rm A}_e(\delta)$ (see Eq.~(\ref{f1})) and $\Sigma^{\rm P}_e (\delta)$ (see Eq.~(\ref{pe})) appear whenever there exists a $m$ such that $\delta \to \varepsilon_m - \varepsilon_e$ and graphically correspond to poles of the Green function crossing the imaginary axis in Fig.~\ref{contour} as $\delta$ is varied.
The choice of convention for $\Theta(0)$ must be made so that $\Sigma^{\rm c}_e (\delta)$ remains continuous in these situations.
To proceed, we begin by observing that the sum of the discontinuous terms in Eq.~(\ref{pe}) and Eq.~(\ref{dhe}) (those for which $n=m$) when $\delta \to (\varepsilon_m - \varepsilon_e)^\pm$ is 
\begin{align*}
- & \frac{1}{2} \bra{em^*} \hat v_C^{1/2} \big( \hat \epsilon^{-1}(0) - \hat 1 \big) \hat v_C^{1/2} \ket{m^*e} & &{\rm if} \ket{m} \in \{\ket{v}\}, \\
+ & \frac{1}{2} \bra{em^*} \hat v_C^{1/2} \big( \hat \epsilon^{-1}(0) - \hat 1 \big) \hat v_C^{1/2} \ket{m^*e} & &{\rm if} \ket{m} \in \{\ket{c}\}. 
\end{align*}
The appropriate choice of convention for $\Theta(0)$ is therefore 
\begin{equation}
\label{theta}
\Theta(x) = 
\begin{cases}
1 & {\rm if} \quad x>0, \\
1/2 & {\rm if} \quad x=0, \\
0 & {\rm if} \quad x<0.
\end{cases}
\end{equation}
Graphically, this convention can be seen as including half of the pole of the Green function inside the contour when it lies precisely on the imaginary axis.
Also, since $\Theta(x)$ takes eigenenergies in argument (in contrast to band indices) due to the form of Eq.~\eqref{green}, degenerate states are uniformly treated by construction and need no special treatment in the present formalism.
Furthermore, in a future generalization of the present implementation to extended systems, the use of $k$ point grids to sample the Brillouin zone will automatically discretize the continuous spectrum of eigenvalues, so that the present treatment of the discontinuities will still apply.
}.

For the highest occupied molecular orbital (HOMO) of silane, choosing $f(\omega) = 1$  allows to converge $\Sigma^{\rm c}_e(\bra{e} \hat \Sigma^{\rm x} - \hat V^{\rm xc} \ket{e})$~\footnote{
The choice $\delta = \bra{e} \hat \Sigma^{\rm x} - \hat V^{\rm xc} \ket{e}$ is motivated by the fact that it is the value of $\delta$ closest to $\Delta \varepsilon_e$ among those we use for our $G_0W_0$ simulations. 
}
to 1~meV with only eight frequency samplings of the integrand of Eq.~(\ref{e_c}) in contrast to 24 when $f(\omega)$ is chosen to be $0$. 

The only nontrivial model function used in previous works was of Gaussian form~\cite{Anisimov:2000ub}. 
This choice yields
\begin{equation*}
\begin{aligned}
f(\omega,\alpha) = & e^{-\omega^2/\alpha^2}, \\
F_{ne}(\delta,\alpha) = & \frac{1}{2} \, {\rm sgn}(\omega_{ne}) e^{\omega_{ne}^2/\alpha^2} \, {\rm erfc} (|\omega_{ne}/\alpha|), \\
\Sigma^{\rm A}_e(\delta,\alpha) =& \frac{1}{2} \sum_{l,l'} \bra{l} \big( \hat \epsilon^{-1}(0) - \hat 1 \big) \ket{l'} \\
& \times \bra{l^*} \hat v_C^{1/2} \hat \Phi_e^\dagger \, {\rm sgn}(\hat H - \varepsilon_{e}-\delta) e^{(\hat H - \varepsilon_{e}-\delta)^2/\alpha^2} \\
& \times {\rm erfc} (|(\hat H - \varepsilon_{e}-\delta)/\alpha|) \hat \Phi_e \hat v_C^{1/2} \ket{l'^*}, \\
\end{aligned}
\end{equation*}
where $\alpha$ is a model parameter that characterizes the width of the Gaussian. 
However, the Taylor expansions required to calculate explicit values for the functions of the Hamiltonian $\hat H$ make this choice numerically too cumbersome to be practical in $G_0W_0$ implementations where summations over conduction states are avoided, like the present method. 
Therefore, in this work, we explore a novel choice of model function, which is both compatible with the elimination of summations over conduction states and physically motivated. 
The chosen form for $f(\omega)$ is a Lorentzian,
\begin{equation}
\label{he}
\begin{aligned}
f(\omega,\alpha) = & \frac{\alpha^2}{\omega^2+\alpha^2}, \qquad (\alpha > 0), \\
F_{ne}(\delta,\alpha) = & \frac{\pi}{2} \frac{\alpha}{\omega_{ne} + \alpha \, {\rm sgn}(\omega_{ne}) }, \\
\Sigma^{\rm A}_e(\delta,\alpha) = & \frac{1}{2} \sum_{l,l'} \bra{l} \big( \hat \epsilon^{-1}(0) - \hat 1 \big) \ket{l'} \\
& \times \big( \bra{l^*} \hat v_C^{1/2} \hat \Phi_e^\dagger \hat {\mathcal Q}_e(\delta) \ket{h_{e,l'}^+(\delta,\alpha)} \\
& + \bra{l^*} \hat v_C^{1/2} \hat \Phi_e^\dagger \hat {\mathcal P}_e(\delta) \ket{h_{e,l'}^-(\delta,\alpha)} \big), \\
\end{aligned}
\end{equation}
where 
\begin{equation}
\label{khe}
\begin{split}
(\hat H - \varepsilon_e-\delta + \alpha) \ket{h_{e,l'}^+(\delta,\alpha)} \equiv & \alpha \hat {\mathcal Q}_e(\delta) \hat \Phi_e \hat v_C^{1/2} \ket{l'^*},\\
(\hat H - \varepsilon_e-\delta - \alpha) \ket{h_{e,l'}^-(\delta,\alpha)} \equiv & \alpha \hat {\mathcal P}_e(\delta) \hat \Phi_e \hat v_C^{1/2} \ket{l'^*},
\end{split}
\end{equation}
and where no complicated functions of $\hat H$ are involved.
Choosing $\alpha=1.0$~Ha in this form of model function allows to converge $\Sigma^{\rm c}_e(\bra{e} \Sigma^{\rm x} - V^{\rm xc} \ket{e})$ for the highest occupied molecular orbital (HOMO) of silane to 1~meV with only four frequency samplings of the integrand of Eq.~(\ref{e_c}) in contrast to eight when $f(\omega) = 1$.

Also, as it can be seen from the following expression, 
\begin{equation}
\begin{split}
\hat \epsilon^{-1}(i \omega)-1 \approx & \frac{\alpha^2}{\omega^2+\alpha^2} (\hat \epsilon^{-1}(0)-1), \\
= & \frac{\alpha}{2} \Bigg( \frac{1}{\alpha+i\omega} + \frac{1}{\alpha-i\omega} \Bigg) (\hat \epsilon^{-1}(0)-1),
\end{split}
\end{equation}
approximating the dynamical character of the inverse dielectric matrix $\hat \epsilon^{-1}(i\omega)-1$ by a Lorentzian is equivalent to replacing all its poles on the positive real axis by a single one.
This choice of model function can therefore be physically interpreted as a scalar version of the plasmon pole model.
It is interesting to note that such a model function has the correct high frequency behavior 
\begin{equation}
\lim_{\omega \to \infty} \frac{\alpha^2}{\omega^2+\alpha^2} \propto \frac{1}{\omega^2} \propto \lim_{\omega \to \infty} \hat \epsilon^{-1}(i\omega)-1 
\end{equation}
while this is not the case for the Gaussian model. 
A generalization of this model to the level of conventional plasmon pole approximations is currently under way. 


\section{An efficient basis $\{\ket{l}\}$ for $\hat \epsilon^{-1}(i\omega)$}
\label{Lanczos}

Eliminating the sums over conduction states that were present in Eqs.~(\ref{e_c})-(\ref{dhe}) required us to introduce a complete basis $\{\ket{l}\}$ in the terms $\sigma^{\rm N}_e(i\omega,\delta)$, $\sigma^{\rm N0}_e(i\omega,\delta,\alpha)$ and $\Sigma^{\rm A}_e(\delta,\alpha)$.
\begin{widetext}
These terms can be rewritten in the form of a trace in the basis $\{\ket{l}\}$, starting from Eqs.~(\ref{ge}), (\ref{kge}), (\ref{g0e}), (\ref{he}), and (\ref{khe}):
\begin{equation}
\label{traces}
\begin{aligned}
\sigma^{\rm N}_e(i\omega,\delta) =& \sum_{l} \bra{l} \big( \hat \epsilon^{-1}(i \omega) - \hat 1 \big) \hat v_C^{1/2} \hat \Phi_e \frac{\hat H^* - \varepsilon_e-\delta}{\omega^2 + (\hat H^* - \varepsilon_e-\delta)^2} \hat \Phi_e^\dagger \hat v_C^{1/2} \ket{l} & &\equiv \sum_{l} \bra{l} \hat \sigma^{\rm N}_e(i\omega,\delta) \ket{l}, \\
\sigma^{\rm N0}_e(i\omega,\delta,\alpha) =& \sum_{l} \bra{l} \big( \hat \epsilon^{-1}(0) - \hat 1 \big) \hat v_C^{1/2} \hat \Phi_e \frac{\hat H^* - \varepsilon_e-\delta}{\omega^2 + (\hat H^* - \varepsilon_e-\delta)^2} \hat \Phi_e^\dagger \hat v_C^{1/2} \ket{l} \frac{\alpha^2}{\omega^2+\alpha^2} & & \equiv \sum_{l} \bra{l} \hat \sigma^{\rm N0}_e(i\omega,\delta) \ket{l} \frac{\alpha^2}{\omega^2+\alpha^2} , \\ 
\Sigma^{\rm A}_e(\delta,\alpha) =& \frac{1}{2} \sum_{l} \bra{l} \big( \hat \epsilon^{-1}(0) - \hat 1 \big) \hat v_C^{1/2} \hat \Phi_e \hat {\mathcal Q}_e^* \frac{\alpha}{\hat H^* - \varepsilon_e-\delta + \alpha} \hat {\mathcal Q}_e^* \hat \Phi_e^\dagger \hat v_C^{1/2} \ket{l} & & \\
& + \frac{1}{2} \sum_{l} \bra{l} \big( \hat \epsilon^{-1}(0) - \hat 1 \big) \hat v_C^{1/2} \hat \Phi_e \hat {\mathcal P}_e^* \frac{\alpha}{\hat H^* - \varepsilon_e-\delta - \alpha} \hat {\mathcal P}_e^* \hat \Phi_e^\dagger \hat v_C^{1/2} \ket{l} & & \equiv \sum_{l} \bra{l} \hat \Sigma^{\rm A}_e(\delta,\alpha) \ket{l}.
\end{aligned}
\end{equation}
\end{widetext}
The smallest orthonormal basis $\{\ket{l}\}$ such that the above traces are converged must contain the subspace associated with the highest eigenvalues of $\hat \sigma^{\rm N}_e(i\omega,\delta)$, $\hat \sigma^{\rm N0}_e(i\omega,\delta)$, and $\hat \Sigma^{\rm A}_e(\delta,\alpha)$. 
This subspace corresponds qualitatively to the intersection of the subspaces associated with the highest eigenvalues (in absolute value) of $\hat v_C^{1/2}$, $\hat \Phi_e$, $(\hat H^* - \varepsilon_e-\delta)^{-1}$, and $\hat \epsilon^{-1}(i\omega) - \hat 1$.
Since $\hat v_C^{1/2}$ is diagonal in the basis of plane-waves, $\hat \Phi_e$ is diagonal in real space and $(\hat H^* - \varepsilon_e-\delta)^{-1}$ is diagonal in the basis of complex-conjugated DFT states $\ket{n^*}$, the subspace generated by the eigenvectors associated with their highest eigenvalues is readily available in the present formalism. 
However, $\hat \epsilon^{-1}(i\omega) - \hat 1$ is not known in any basis yet and, therefore, the subspace associated to its highest eigenvalues remains to be found.

In conventional plane-waves implementations of $G_0W_0$, the dielectric matrix $\hat \epsilon$ is obtained in a plane-wave basis [using Eq.~(\ref{P}) and~(\ref{epsilon})] and directly inverted.
This inversion then becomes a bottleneck, since the size of the basis in which wave functions are expanded is quite large ($\sim115,000$ plane-waves for a molecule as simple as silane). 
Usually, this bottleneck is mitigated by expressing the dielectric matrix in a plane-wave basis smaller than the one used for the wave functions~\cite{Hybertsen86}. 
However, this practice leads to increased uncertainties~\cite{Shih:2010ft} and the calculation size remains limited by this factor. 
It is therefore desirable to obtain a basis spanning the subspace associated with the highest eigenvalues of $\hat \epsilon^{-1}(i\omega) - \hat 1$ directly, without explicitly expressing $\hat \epsilon$ in a plane-wave basis. 

It is useful at this stage to have some insight into the spectrum of $\hat \epsilon(i\omega) - \hat 1$ for $\omega \in \mathbb{R}$.
From Eqs.~(\ref{epsilon}) and (\ref{P}), this operator can be written as 
\begin{equation}
\label{epsilon_nonp}
\hat \epsilon(i\omega) - \hat 1 = 4 \sum_{cv} \hat v_C^{1/2} \ket{c^* v} \frac{\varepsilon_c - \varepsilon_v}{(\varepsilon_c - \varepsilon_v)^2 + \omega^2} \bra{v c^*} \hat v_C^{1/2},
\end{equation}
where $\big(\varepsilon_c - \varepsilon_v\big)/\big((\varepsilon_c - \varepsilon_v)^2 + \omega^2\big) > 0 \ \forall \ \omega \in \mathbb{R}$ and where $\hat v_C^{1/2}$ has only positives eigenvalues. 
It results that $\hat \epsilon(i\omega) - \hat 1$ has only positives eigenvalues. 
In particular, for isolated systems, its eigenvalue spectrum is formed by a few large discrete eigenvalues associated with transitions from valence states to bound conduction states and a continuous spectrum of smaller eigenvalues associated to transitions from valence states to a continuum of conduction states, with an integrable divergence at the origin~\cite{Wilson:2009p5717}.

Moreover, since the eigenvalues of $\hat \epsilon(i\omega) - \hat 1$ are in the range $[0,+\infty[$, those of $1 - \hat \epsilon^{-1}$ are located in the range $[0,1[$.
Also, if the eigenvalues of both operators are sorted in decreasing order, the corresponding eigenvalues will occupy the same rank.
Consequently, the eigenvalues of $1 - \hat \epsilon^{-1}$ that contribute most to $\Sigma^{\rm c}_e (\delta)$ correspond to the largest eigenvalues of $\hat \epsilon - \hat 1$.
However, in contrast to $1 - \hat \epsilon^{-1}$, it is possible to apply  $\hat \epsilon - \hat 1$ on an arbitrary vector without having to explicitly construct its matrix representation, using Eqs.~(\ref{epsilon}), (\ref{P1}) and (\ref{P2}).

Therefore, to approximately find the subspace associated to its largest eigenvalues, one can apply  $\hat \epsilon - \hat 1$ repeatedly on some random vector $\ket{\psi}$ to construct a Krylov subspace~\cite{Golub}:
\begin{equation}
\label{krylov}
\{ \ket{\psi}, (\hat \epsilon - \hat 1)\ket{\psi}, (\hat \epsilon - \hat 1)^2 \ket{\psi}, ..., (\hat \epsilon - \hat 1)^N \ket{\psi} \}.
\end{equation}
Applying $\hat \epsilon - \hat 1$ on a random vector will cause the directions associated to the largest eigenvalues to grow faster than the others with respect to the number of applications. 
Thus, orthonormalizing the vectors of Eq.~(\ref{krylov}) would yield a basis approximately generating the desired subspace. 

In practice, our implementation uses the vector $\ket{\psi} = \hat v_C^{1/2} \ket{e^*e}$ as a starting point, but we observed that using any other vector (or all other vectors) of the form $\ket{\psi} = \hat v_C^{1/2} \ket{v^*e}$ does not change significantly the number of dimensions $N$ required to achieve convergence. 
This also implies that the basis generated from $\ket{\psi} = \hat v_C^{1/2} \ket{e^*e}$ can be used to obtain $G_0W_0$ corrections for all desired DFT eigenstates and not only $\ket{e}$.

Also, in practice, we use the Lanczos procedure~\cite{Golub} to obtain an orthogonal basis that spans the subspace of Eq.~(\ref{krylov}) and tridiagonalizes $\hat \epsilon - \hat 1$.
This procedure also provides the associated matrix elements and its cost is only marginally higher than the successive applications of $\hat \epsilon - \hat 1$ on $\ket{\psi}$.

Theoretically, the Lanczos procedure should not require explicit orthogonalization of each basis vector with all previously generated ones.  
However, in practice, the vectors generated by a direct implementation of the Lanczos procedure rapidly loose their orthogonality with the number of steps due to numerical error~\cite{Golub}.
For example, in our implementation, orthogonality is typically lost in about ten steps, while 100's of steps are required to achieve convergence. 
Therefore, in our implementation of the Lanczos procedure, we added a Gram-Schmidt orthogonalization of each residual vector with respect to all previously generated Lanczos vectors, so that orthogonality is achieved to machine precision.
A fully converged calculation of $\Sigma^{\rm c}_e(\delta)$ for the HOMO of silane and an identical one except for the omission of the orthogonalization of Lanczos vectors have shown a difference in computation time of less than $0.05$~\%.
Thus, the necessity of introducing this orthogonalization has no impact on the performance of the present $G_0W_0$ implementation. 

At this stage, it is possible to invert the dielectric matrix $\hat \epsilon$ at a very low computational cost, since the Lanczos basis is much smaller then the plane-waves basis and since a tridiagonal matrix can be inverted at a cost $\propto N^2$, in contrast to $N^3$ for a full matrix.
We thus approximately obtain $1 - \hat \epsilon^{-1}$ in the subspace of its $N$ largest eigenvalues, which contribute most per dimension to the quasiparticle energy $\Sigma^{\rm c}_e (\delta)$. 
An alternative way to obtain $\hat \epsilon$ in approximately the same subspace is to iteratively diagonalize it~\cite{Wilson:2009p5717,Wilson:2008p6437}, as was implemented in another plane-waves $G_0W_0$ code~\cite{Pham:2013gs}.
However, the latter scheme costs about ten applications of $\hat \epsilon$ per dimension while our scheme costs a single application per dimension. 
This efficiency gain is made possible by the fact that adding an extra dimension to the Lanczos basis does not only converge the traces of Eq.~(\ref{traces}) by adding a term to the sum.
It also increases the agreement between all the eigenvalues of the tridiagonal matrix and the corresponding eigenvalues of the exact operator.
Moreover, according to the Kaige-Pangel theory~\cite{Golub}, the quickest convergence is achieved for the eigenvalues most separated from the others, which are the largest ones in the case of the dielectric matrix, e.g., the ones contributing most to the final result $\Sigma^{\rm c}_e(\delta)$. 
Thus constructing the dielectric operator in a whole subspace at once using the Lanczos algorithm allows for a substantial efficiency gain with respect to iterative diagonalization techniques, mostly due to the fact that the latter cannot use the information obtained in the construction of an eigenvector/eigenvalue pair for the refinement of another pair.

There remains only one step to obtain a basis $\{\ket{l}\}$ that approximately spans the smallest possible subspace where the traces of Eq.~(\ref{traces}) are converged.
It is to intersect the Lanczos basis obtained for $\hat \epsilon$ with the subspaces associated with the largest eigenvalues of the operators $\hat v_C^{1/2}$, $\hat \Phi_e$ and $(\hat H^* - \varepsilon_e-\delta)^{-1}$.
However, the cost of calculating the traces in Eq.~(\ref{traces}) is of order $N^2 \ln(N)$ for a plane-wave implementation, which is lower than the $N^3$ order associated to the projection of the Lanczos vectors on the relevant subspaces.  
Moreover, leaving extra dimensions in the basis $\{\ket{l}\}$ just increases the precision of the calculated trace.
Therefore, in plane-waves implementations, the most efficient choice of basis $\{\ket{l}\}$ for the calculation of the traces in Eq.~(\ref{traces}) is to pick the smallest of the four aforementioned subspaces, that is, the Lanczos basis for the dielectric matrix $\hat \epsilon$.
As an example, calculations of $\Sigma^{\rm c}_e(\delta)$ for the HOMO of silane require a Lanczos basis of dimension $\sim 500$ to be converged to $\sim 20$~meV, while conventional plane-waves $G_0W_0$ calculations require a dielectric matrix of dimension $\sim 15,000$ to achieve a similar convergence.  

\section{Dielectric model}
\label{model}
We mentioned in Sec.~\ref{Lanczos} that, when the Lanczos basis $\{\ket{l}\}$ of the dielectric operator $\hat \epsilon - \hat 1$ is iteratively constructed, the fastest converging eigenvalues are the largest ones. 
Therefore, as the construction of $\{\ket{l}\}$ progresses, an increasing proportion of the computational work becomes aimed at sampling the integrable divergence in the spectrum of eigenvalues of $1 - \hat \epsilon^{-1}$, while the large eigenvalues are already converged.
We have also mentioned in Sec.~\ref{Lanczos} that the small, continuous eigenvalues making up this integrable divergence are associated with transitions from valence bands to free conduction states. 
Thus, at some point in the construction of $\{\ket{l}\}$, when applying $\hat \epsilon - \hat 1$ to the current basis vector $\ket{l}$, the largest terms in the sum of Eq.~(\ref{epsilon_nonp}) will be associated to eigenenergies $\varepsilon_c$ of free conduction states, dominated by kinetic energy.
We will then have $\hat H \ket{c} = \varepsilon_c \ket{c} \approx \hat T \ket{c}$, where $\hat T$ is the kinetic energy operator. 
Therefore, substituting $\hat H \to \hat T$ in Eqs.~(\ref{P2a})-(\ref{P2+}) should become an accurate approximation at some point in the construction of $\{\ket{l}\}$. 

Since $\hat T$ is diagonal in the plane-wave basis, in contrast to $\hat H$, the conversion of Eq.~(\ref{P2a}) into a linear equation problem [Eq.~(\ref{P2})] is no longer required. 
Indeed, the former equation can then be directly solved with a single application of $\hat T^{-1}$ on a vector at a cost $\propto N$, where $N$ is the number of plane-waves in the basis in which the wave functions are expressed. 
In contrast, the solution of Eqs.~(\ref{P2}) or (\ref{P2+}) using SQMR requires typically $\sim 15$ application of the Hamiltonian $\hat H$ on a vector at a cost $\propto N {\rm ln}(N)$.
Since these successive applications of $\hat H$ are the bottleneck in the construction of $\{\ket{l}\}$, it becomes interesting to use the approximation $\hat H \to \hat T$ in Eq.~(\ref{P2a}) as soon as it becomes accurate. 

To implement this idea, we first define the approximate dielectric operator from Eqs.~(\ref{epsilon}), (\ref{P1}) and (\ref{P2a}), where we substitute $\hat H \to \hat T$:
\begin{equation}
\label{model_eq}
\begin{aligned}
\hat {\tilde \epsilon}(\omega) \equiv & 1 - \hat v_C^{1/2} \hat {\tilde P}(\omega) \hat v_C^{1/2}, \\
\hat {\tilde P} (\omega) \ket{\psi_j} \equiv & - 2 \sum_v \ket{v} \Big( \ket{\tilde f^*_{jv-}(\omega)} + \ket{\tilde f^*_{jv+}(\omega)} \Big), \\
\ket{\tilde f_{jv\pm}(\omega)} \equiv & \hat {\mathcal P}_c \sum_G \ket{G} \frac{1}{G^2/2 - \varepsilon_v \pm \omega} \bra{G} \hat {\mathcal P}_c \ket{v \psi_j^*},
\end{aligned}
\end{equation}
where $\ket{G}$ is a plane-wave of the basis used to express the wave functions and $G$ is the corresponding wavevector. 
Then, we add and subtract the model operator $\hat {\tilde \epsilon}^{-1}$ from the exact operator $\hat \epsilon^{-1}$, 
\begin{equation}
\label{subs_epsilon}
\hat \epsilon^{-1} - 1 = (\hat \epsilon^{-1} - \hat {\tilde \epsilon}^{-1}) + (\hat {\tilde \epsilon}^{-1} - 1)
\end{equation}
in $\sigma^{\rm N}_e$, $\sigma^{\rm N0}_e$, and $\Sigma^{\rm A}_e$ as expressed in Eq.~(\ref{traces}).
Since $\hat {\tilde \epsilon}^{-1}$ should accurately describes the integrable divergence in the spectrum of eigenvalues of $\hat \epsilon^{-1}$ near $1$, the operator $(\hat \epsilon^{-1} - \hat {\tilde \epsilon}^{-1})$ should be devoid of such a divergence. 
Therefore, the calculation of its trace should require a smaller Lanczos basis than the $(\hat \epsilon^{-1} - 1)$ operator or the $(\hat {\tilde \epsilon}^{-1} - 1)$ operator.

We exploit this by splitting each trace of Eq.~(\ref{traces}) in two others.
In the first traces, we substitute $(\hat \epsilon^{-1} - 1)$ by $(\hat \epsilon^{-1} - \hat {\tilde \epsilon}^{-1})$ and can thus use a smaller Lanczos basis.
In the second traces, we substitute $(\hat \epsilon^{-1} - 1)$ by $(\hat {\tilde \epsilon}^{-1} - 1)$ and the basis size will remain similar.
Thus, we use the exact dielectric operator $\hat \epsilon - 1$ to generate the first basis $\{\ket{l}\}$ at a reduced cost, thanks to its smaller size.
Then, we use the approximate dielectric operator $\hat {\tilde \epsilon} - 1$ to generate the second basis $\{\ket{\tilde l}\}$, also at a reduced cost, since the operator is simpler to apply.

Once those bases are available, we split each of $\sigma^{\rm N}_e$, $\sigma^{\rm N0}_e$, and $\Sigma^{\rm A}_e$ as expressed in Eqs.~(\ref{ge}), (\ref{g0e}) and (\ref{he}) in a sum of two contributions.
In the first contributions, we substitute $(\hat \epsilon^{-1} - 1)$ by $(\hat \epsilon^{-1} - \hat {\tilde \epsilon}^{-1})$ and evaluate the resulting expression using the basis $\{\ket{l}\}$.
In the second contributions, we substitute $(\hat \epsilon^{-1} - 1)$ by $(\hat {\tilde \epsilon}^{-1} - 1)$ and evaluate the resulting expression using the basis $\{\ket{\tilde l}\}$.
Then, we sum the two results and obtain $\sigma^{\rm N}_e$, $\sigma^{\rm N0}_e$ and $\Sigma^{\rm A}_e$ at a reduced computational cost. 

It is interesting to note that, since the dielectric model is subtracted and added to the exact dielectric operator, the scheme described in this section does not introduce new approximations in the $G_0W_0$ formalism, provided that the size of the Lanczos bases $\{\ket{l}\}$ and $\{\ket{\tilde l}\}$ are sufficient to obtain converged results. 

\section{Calculating the dielectric matrix and the integrand at different frequencies}
\label{recyclage}
In principle, the integration over frequencies $\omega$ in Eq.~(\ref{e_c}) could be carried out by using Eqs.~(\ref{ge}), (\ref{kge}), (\ref{g0e}) and by building a new Lanczos basis $\{\ket{l}\}$ at each different value of $\omega$ (we consider the case where the dielectric model is not used for simplicity). 
However, in practice, the Lanczos bases for $\hat \epsilon(i\omega)$ at different imaginary $i\omega$ span subspaces that do not differ significantly.
This can be understood from Eq.~(\ref{P}) for the polarizability, which can be rewritten as
\begin{equation}
P(i\omega) = -4 \sum_{cv} \ket{c^*v} \frac{\varepsilon_c - \varepsilon_v}{\omega^2 + (\varepsilon_c - \varepsilon_v)^2} \bra{vc^*}.
\end{equation}
When $i\omega$ is displaced along the positive direction of the imaginary axis, each term of the sum decreases monotonically.
In contrast, making $i\omega$ real and displacing it along the real axis in Eq.~(\ref{P}) would cause strong changes in the terms of the summation, since they each contain one pole located on the real axis. 
Thus, we found that constructing a basis for the static dielectric matrix $\hat \epsilon(0)$ and using it to express the dynamic dielectric matrix $\hat \epsilon(i\omega)$ at all other frequencies $i\omega \in [0,i\infty]$ is a sound approximation, in agreement with previous work~\cite{Pham:2013gs,vonderLinden:1988ca}. 


Still, a new Lanczos basis must be constructed to calculate the dielectric matrix at each real frequency required by Eq.~(\ref{pe}).
However, in contrast to the other contributions to $\Sigma^{\rm c}_e(\delta)$ as expressed in Eqs.~(\ref{e_c}) and (\ref{traces}), only a single matrix element of $\hat \epsilon^{-1} - \hat 1$ per frequency is required in Eq.~(\ref{pe}) instead of some related trace.
This causes Eq.~(\ref{pe}) to converge dramatically faster than Eq.~(\ref{traces}) with respect to the size of the Lanczos basis, provided the seed vector is chosen appropriately.
Indeed, to calculate a matrix element of the inverse dielectric matrix $\bra{ep^*} \hat v_C^{1/2} \big( \hat \epsilon^{-1}(\omega_{pe}) - \hat 1 \big) \hat v_C^{1/2} \ket{p^*e}$, where $\ket{p}$ is the state generating the pole whose residual is being calculated, knowledge of the dielectric matrix in a subspace formed by the eigenvectors that both correspond to its largest eigenvalues and overlap substantially with the vector $\hat v_C^{1/2} \ket{p^*e}$ is required. 
The latter will automatically be satisfied if the seed vector is chosen to be $\hat v_C^{1/2} \ket{p^*e}$ and the former is a feature of the Lanczos procedure. 
Thus, with this choice of seed vector, we found that four Lanczos iterations converge $\Sigma^{\rm P}_e(\delta)$ to 1~meV for all the systems studied. 
Together with the small number of terms involved by Eq.~(\ref{pe}) when only states close to the band gap are corrected, the preceding observation keeps the computational time spent on $\Sigma^{\rm P}_e(\delta)$ small with respect to the remainder of $\Sigma^{\rm c}_e(\delta)$.
Thus, the necessity of building a separate Lanczos basis for each real frequencies present in Eq.~(\ref{pe}) has a small impact on the performance of the implementation. 

Also, keeping the basis in which $\hat \epsilon$ is expressed fixed for all $i\omega$ enables some tricks to speed up the calculation. 
For instance, we adopt the shift Lanczos technique to solve Eq.~(\ref{kge}) simultaneously at all frequencies~\cite{Pham:2013gs, Nguyen:2012iy} (a few tens of iterations is typically enough to converge $\Sigma^{\rm c}_e$ to 5~meV).
This technique requires that the linear equation have the general form $(\hat M + \hat I \omega) \ket{x(\omega)}  = \ket{b}$ where both the operator $\hat M$ and the right-hand side $\ket{b}$ are independent of $\omega$. 
However, here $\ket{b} = ( \hat H - \varepsilon_e-\delta ) \hat \Phi_e \hat v_C^{1/2} \ket{l^*}$.
Therefore, if $\{\ket{l}\}$ was dependent on $\omega$, Eq.~(\ref{kge}) would have to be solved individually for each frequency $\omega$.
Keeping the basis $\{\ket{l}\}$ fixed is thus required to allow Eq.~(\ref{kge}) to be solved at all frequencies simultaneously. 

It would also be possible to use the same technique to solve Eq.~(\ref{P2+}) (where we substitute $\ket{\psi_j} \to \hat v^{1/2}_C \ket{l}$ and $j \to l$) simultaneously at all frequencies, at the cost of applying the Hamiltonian $\hat H$ a few tens of times per dimension of the dielectric matrix~\cite{Pham:2013gs,Nguyen:2012iy}. 
However, it is possible to avoid any iterative solution of Eq.~(\ref{P2+}) in a plane-wave basis beside those that were already solved in the construction of the static dielectric matrix $\hat \epsilon(0)$. 
To do this, we construct one basis per valence state $v$, $\{\ket{\gamma_{i,v}}\}$, much smaller than the plane-wave basis, in which the Hamiltonian $\hat H$ and the right-hand side of Eq.~(\ref{P2+}) are expressed, which allows to solve the latter equation by direct inversion of the resulting matrix at a negligible computational cost.
To suitably choose the basis $\{\ket{\gamma_{i,v}}\}$, it is useful to rewrite the quantity we wish to calculate with the solution of Eq.~(\ref{P2+}), that is, $\bra{l} \hat \epsilon(i\omega) - \hat 1 \ket{l'}$, from Eqs.~(\ref{epsilon}), (\ref{P1}), (\ref{P2+}), and (\ref{fv})
\begin{equation}
\label{recy_eq}
\begin{split}
\bra{l} \hat \epsilon(i\omega) - \hat 1 \ket{l'} = & 4 \sum_v \bra{l} \hat v_C^{1/2} \hat \Phi_v \hat {\mathcal P}_c^* \\
& \times \frac{\hat H^* - \varepsilon_v}{(\hat H^* - \varepsilon_v)^2 + \omega^2} \hat {\mathcal P}_c^* \hat \Phi_v^\dagger \hat v_C^{1/2} \ket{l'}, \\
= & 4 \sum_v \bra{b_{l,v}} \hat A_v^{-1}(i\omega) \ket{b_{l',v}}, \\
= & 4 \sum_v \braket{b_{l,v} | x_{l',v}(i\omega)},
\end{split}
\end{equation}
where $\ket{b_{l,v}} \equiv \hat {\mathcal P}_c^* \hat \Phi_v^\dagger \hat v_C^{1/2} \ket{l}$, $\hat A_v(i\omega)  \equiv \big( (\hat H^* - \varepsilon_v)^2 + \omega^2 \big) / (\hat H^* - \varepsilon_v)$ and $\ket{x_{l,v}(i\omega)} \equiv \hat A_v^{-1}(i\omega) \ket{b_{l,v}}$.
The optimal basis to pick as $\{\ket{\gamma_{i,v}}\}$ would be the one spanning the same subspace as the eigenvectors $\ket{\lambda_v(i\omega)}$ of $\hat A_v(i\omega)$ that contribute most to the desired quantity ($\bra{l} \hat \epsilon(i\omega) - \hat 1 \ket{l'} = 4 \sum_{v,\lambda} \braket{b_{l,v} | \lambda_v(i\omega)} 1/\lambda_v(i\omega) \braket{\lambda_v(i\omega) | b_{l,v}}$), that is, those that are both associated to small eigenvalues $\lambda_v(i\omega)$ of $\hat A_v(i\omega)$ and overlapping substantially with the states $\ket{b_{l,v}}$.
The solutions $\ket{x_{l,v}(i\omega)} = \sum_\lambda \ket{\lambda_v(i\omega)} 1/\lambda_v(i\omega) \braket{\lambda_v(i\omega) | b_{l,v}}$ are naturally dominated by these directions, so that $\{\ket{x_{l,v}(i\omega)}\}$ should be a proper choice of basis in which to express Eq.~(\ref{P2+}).

There only remains to pick the frequency $i\omega$ at which the solutions $\{\ket{x_{l,v}(i\omega)}\}$ will be used to build the basis.
At this point, it is interesting to note that $\hat A_v (i\omega)$ is simply the shifted conjugated Hamiltonian at the beginning of the domain of integration [$\hat A_v(0) = (\hat H^* - \varepsilon_v)$] and monotonically evolves toward the inverse of this operator as the frequency increases [$\lim_{i\omega \to +i\infty} \hat A_v(i\omega) \to (\hat H^* - \varepsilon_v)^{-1} \times \omega^2$]. 
The two most natural frequencies to select would therefore be $0$ and $i\infty$. 
Since $\lim_{i\omega \to +i\infty} \ket{x_{l,v}(i\omega)} \to (\hat H^* - \varepsilon_v) \times \omega^{-2} \ket{b_{l,v}}$, the second set of solutions is available at a negligible computational cost.
Also, Eq.~(\ref{P2+}) has already been solved at $i\omega=0$ for all $l$ and $v$ in the process of building the static dielectric matrix. 
Therefore, the $\ket{x_{l,v}(0)}$ are available at no additional computational cost. 

The resulting solutions can be made into a basis using singular value decomposition~\cite{Golub} and orthonormalization. 
Finally, $H_{ij,v} = \bra{\gamma_{i,v}} (\hat H - \varepsilon_v) \ket{\gamma_{j,v}}$ and $b_{i,l,v} = 2 \bra{\gamma_{i,v}} \hat {\mathcal P}_c \hat \Phi_v \hat v_C^{1/2} \ket{l^*}$ are computed once and Eq.~(\ref{P2+}) can be solved at all required frequencies by direct inversion of $\hat A_v(i\omega)$ in the $\{\ket{\gamma_{i,v}}\}$ basis at a very low computational cost.
Indeed, calculation times for the $G_0W_0$ correction to the HOMO of silane show only random fluctuations of about 1\% when the number of frequencies considered for the numerical integration of Eq.~(\ref{e_c}) is varied between 1 and 12. 

It is possible to control the accuracy of this choice of basis by adding a set of solutions $\ket{x_{l,v}(i\omega_0)}$ to those we have already selected and observing that the result is negligibly affected.
We have selected for $\omega_0$ a value of $1.0$~Ha, which is approximately the frequency at which the integrand in Eq.~(\ref{e_c}) is maximal for the HOMO of silane. 
The resulting value of $\Sigma^{\rm c}_e(\delta)$ was affected by about 7~meV.
Further testing with values of $0.1$~Ha and $0.01$~Ha for $\omega_0$ has shown similar behavior. 



\section{Theoretical analysis of computational cost}
\label{scaling}
Provided that DFT eigenstates $\ket{n}$ and eigenvalues $\varepsilon_n$ have already been calculated, our implementation starts with the construction of the Lanczos bases $\{\ket{l}\}$ and $\{\ket{\tilde l}\}$ as described in Sec.~\ref{Lanczos} and \ref{model}.
The bottlenecks of this step are the applications of $\hat \epsilon(0)$ on the $N_L$ Lanczos vectors $\ket{l}$ as described by Eqs.~(\ref{epsilon}), (\ref{P1}) and (\ref{P2}) and the applications of $\hat {\tilde \epsilon}(0)$ on the $N_{\tilde L}$ Lanczos vectors $\ket{\tilde l}$ as described by Eq.~(\ref{model_eq}).
By analyzing the first bottleneck, we find that for each of the $N_v$ terms of the sum in Eq.~(\ref{P1}), Eq.~(\ref{P2}) needs to be solved iteratively, which involves $N_{SQMR}$ applications of $\hat H - \varepsilon_v$ on a vector and the same amount of orthogonalization with the valence wave functions. 
Applying the Hamiltonian costs $N_{FFT} \ln(N_{FFT})$ operations in a plane-waves implementation, where $N_{FFT}$ is the number of components describing a wave function in real space, while the orthogonalization costs $N_v N_{PW}$, where $N_{PW}$ is the number of components describing a wave function in reciprocal space.
Thus, we conclude that the worst scaling in this bottleneck comes from the orthogonalization and is $\propto N_L N_v N_{SQMR} N_v N_{PW}$, which is $\propto N^4$, since $N_{SQMR}$ is independent of the system size.
Similarly, the worst scaling in the construction of $\{\ket{\tilde l}\}$ is $\propto N_{\tilde L} N_v N_v N_{PW} \propto N^4$.

It turns out that this $N^4$ scaling is also the worst scaling found in the remainder of the code, as can be expected in general for plane-waves $G_0W_0$ codes~\cite{Aulbur00}. 
We proceed below with an exhaustive list of the operations in our implementation having this scaling.
We limit ourselves to the case where no dielectric model is used, since the scalings involved in the latter are all identical to the corresponding scalings for the exact dielectric operator.

Once the Lanczos basis $\{\ket{l}\}$ is available, our implementation starts the calculations of Eq.~(\ref{e_c}) by the numerical integration of $\sigma^{\rm N}_e(i\omega,\delta) - \sigma^{\rm N0}_e(i\omega,\delta)$, whose integrand can be calculated using Eqs.~(\ref{epsilon}), (\ref{P1}), (\ref{P2+}), (\ref{fv}), (\ref{ge}), (\ref{kge}) and (\ref{g0e}).
The process starts with the construction, for each valence state $v$, of a basis $\{\ket{\gamma_{i,v}}\}$ of dimension $\propto N_L$ for the Sternheimer equation [see Eq.~(\ref{P2+})], as described in Sec.~\ref{recyclage}.
This involves $N_v$ singular value decompositions that scale as $N_L^2 N_{PW}$, which results in a total cost $N_v N_L^2 N_{PW} \propto N^4$.
Then, the projection of Eq.~(\ref{P2+}) in these bases also cost $N_v N_L^2 N_{PW} \propto N^4$.
Finally, the solution of Eq.~(\ref{P2+}) by direct inversion of $(\hat H - \varepsilon_v)^2 - (i\omega)^2$ in these bases scales as $N_\omega N_v N_L^3$.
Since $N_\omega$ is the number of frequencies used for the numerical integration and is not dependent on system size, we obtain $N_\omega N_v N_L^3 \propto N^4$.
All other operations in this section of the code scale, at worst, as $N^3 \ln(N)$.

Next, our implementation calculate $\Sigma^{\rm A}_e(\delta)$ using Eqs.~(\ref{he}) and (\ref{khe}).
However, this part of the code scales as $N^3$ and is not a bottleneck of the calculation.

Finally, our implementation proceeds with the calculation of $\Sigma^{\rm P}_e(\delta)$ using Eqs.~(\ref{pe}), (\ref{epsilon}), (\ref{P1}) and (\ref{P2}) as well as the Lanczos algorithm described in Sec.~\ref{Lanczos}.
However, again, this part of the code scales as $N^3$ and is not a bottleneck of the calculation.


\section{Relation with existing implementations}
\label{relation}
As mentioned in the introduction, the conversion of summations over conduction states into Sternheimer equations is a well established technique both in density functional perturbation theory~\cite{Baroni01,Gonze:1995p6284} and $G_0W_0$.
It was first applied to the latter more than 15 years ago by Ref. \onlinecite{b-24} and has been successfully used in other implementations~\cite{Umari10,Lambert:2013bz,Pham:2013gs} since then.
In all cases, the sums over conduction states present in Eqs. \eqref{green} and \eqref{P} are converted into linear equations analogous to Eqs. \eqref{P2+} and \eqref{kge}.
It is also common to solve these equations at imaginary frequencies, to avoid the poles on the real axis~\cite{Umari10,Lambert:2013bz,Pham:2013gs}.
However, the use of the SQMR algorithm~\cite{Freund:1995vu} allows us to solve these equations at real frequencies, thus enabling the use of the contour deformation technique~\cite{Giantomassi:2011ev,Lundqvist68}, in contrast to the aforementioned implementations, which either use analytic continuation~\cite{Rieger:1999p7438} (Ref. \onlinecite{Pham:2013gs}, Ref. \onlinecite{Umari10} and Ref. \onlinecite{Lambert:2013bz}) or plasmon pole models~\cite{Hybertsen86} (Ref. \onlinecite{b-24}). 
Our implementation is thus distinctively devoid of approximations on the frequency dependence of the self-energy or the dielectric matrix.

It is also common to solve the Sternheimer equations simultaneously at all relevant frequencies using a multi-shift linear equation solver, either Lanczos-based~\cite{CB-13,Walker:2006dq} (present work, Ref. \onlinecite{Pham:2013gs} and Ref. \onlinecite{Umari10}), Frommer's multi-shift solver~\cite{b-29} (Ref. \onlinecite{Lambert:2013bz}), or the Taylor expansion of $(\hat H - \omega)^{-1}$ (Ref. \onlinecite{b-24}).
However, our strategy to reduce the computational cost of the solution of Eq.~\eqref{P2+} below the level of a multi-shift linear equation solver, as presented in the second half of Sec.~\ref{recyclage}, has not previously been described in the literature to our knowledge. 
Indeed, while Ref.~\onlinecite{Umari10} uses a singular value decomposition (SVD)-like procedure on the $\{\ket{b_{l,v}}\}$ vectors of Eq.~\eqref{recy_eq} to reduce the number of solutions $\ket{x_{l,v}(\omega)}$ to be obtained using multi-shift Lanczos, our implementation improves on this procedure by performing the SVD on the (already available) solutions $\{\ket{x_{l,v}(0)},\ket{x_{l,v}(\infty)}\}$ and directly solving Eq.~\eqref{recy_eq} in the resulting basis.
This divides the number of applications of the Hamiltonian involved in the procedure by, approximately, the number of multi-shift Lanczos steps. 

The expression of the dielectric matrix in a Lanczos basis is the core improvement proposed in this article to the $G_0W_0$ method.
Indeed, it should outperform the iterative diagonalization of the dielectric matrix used in Ref.~\onlinecite{Pham:2013gs} and Ref.~\onlinecite{Umari10}, since the latter requires $\sim$10 applications of the dielectric matrix per dimension, while our method requires a single application per dimension, with a comparable number of dimensions for both methods at a given convergence criterion (see Sec.~\ref{Lanczos}).
It also outperforms the self-consistent evaluation of the dielectric matrix (under the form of the screen coulomb potential) proposed in Ref.~\onlinecite{Lambert:2013bz}, since the latter does not outperform the calculation of the dielectric matrix in a plane-wave basis at small system size, while our method does.

Finally, it is common to keep the basis of the dielectric matrix fixed with respect to imaginary frequencies, since it is prerequisite for the use of multi-shift linear equation solvers, as described earlier in this section as well as in Sec.~\ref{recyclage}.
Indeed, this approximation has successfully been used in all implementations that generate an optimal basis to represent the dielectric matrix, that is, Ref.~\onlinecite{Pham:2013gs}, Ref.~\onlinecite{Umari10}, and the present implementation.

\section{Results}
\label{results}
The ABINIT software package~\cite{Gonze:2009aa} is used to produce the DFT eigenstates and eigenvalues used as input in our $G_0W_0$ calculations.
To ease reproducibility, we choose to simulate all molecules with their experimental geometry: silane~\cite{NIST:bench}, thiophene~\cite{pubchem}, benzene~\cite{pubchem}, naphthalene~\cite{pubchem}, anthracene~\cite{pubchem}, tetracene~\cite{pubchem} and C$_{60}$~\cite{GunnarssonBook,DAVID:1991p167}.
We use LDA (Teter Pade parametrization~\cite{Goedecker:1996p9207}) and/or PBE~\cite{PBE} functional for DFT calculations, as specified in each result table.
The corresponding Hartwigsen-Goedeker-Hutter LDA~\cite{ABINIT:site} or PBE~\cite{Krack:2005gf} pseudopotentials are used throughout.  
The reference $G_0W_0$ implementation in Sec.~\ref{performance} is the conventional $G_0W_0$ implementation present in ABINIT. 

For all systems studied, values for the highest occupied molecular orbital (HOMO) are provided. 
However, for the lowest unoccupied molecular orbital (LUMO), we concentrate on molecules where the orbital is bound (anthracene, tetracene, and C$_{60}$).
Indeed, for unbound cases, both $G_0W_0$ and experimental results are scarcer.
Due to this diminished interest, and because plane-wave $G_0W_0$ calculations of molecules are cumbersome even with our implementation, we only provide results for the computationally simplest of these cases, the LUMO of silane.

Since this study focuses on isolated systems, a spherical truncation of the Coulomb potential~\cite{Spencer:2008gz} is used in both the current implementation and the reference implementation of $G_0W_0$.
Its radius is set to half the size of the cubic unit cell used to simulate the molecule.
Thus converging the $G_0W_0$ correction [see Eq.~\eqref{exc}] with respect to the size of the unit cell ensures both that the wave functions of a given molecule are coupled together over their full spatial extent and that the wave functions belonging to different periodic replicas of the molecule do not interact.
However, the spherical truncation of the Coulomb potential is not implemented in the DFT part of ABINIT, so that this strategy cannot be applied to the DFT eigenenergies. 
Moreover, in periodic simulations, the latter are difficult to position with respect to vacuum.
To simultaneously correct them for the spurious Coulomb interactions between periodic replicas of the molecule and position them with respect to the energy level of the vacuum, we also run DFT calculations using the BigDFT project~\cite{Genovese:2008hr}.
This code uses a localized basis set (Daubechies wavelets~\cite{Daubechies:1992wb}) and a Poisson solver with free boundary conditions~\cite{Genovese:2006eb}, which allow for nonperiodic simulations, free of spurious Coulombic interactions, and a physical determination of the energy level of the vacuum.
A spacing of 0.4~bohr for the wavelet grid is used to converge the eigenenergies to a few meV. 
The difference between the eigenenergies obtained using BigDFT and ABINIT can then be added to the $G_0W_0$ eigenenergies to ensure that they are correctly positioned with respect to the vacuum and devoid of spurious Coulombic interactions, provided that the $G_0W_0$ correction is converged with respect to the unit cell size.

Also, Gauss-Legendre quadratures with eight points are used to integrate Eq.~(\ref{e_c}) (or Eq.~(\ref{e_c3}) for the reference $G_0W_0$ implementation). 
DFT eigenvalues and eigenstates are converged until the squared residual of the wave functions $|(\hat H - \varepsilon_n) \ket{n}|^2$ is less than $10^{-12}$~Ha$^2$. 
The solutions to Eqs.~(\ref{P2}) or (\ref{P2+}) are iteratively refined until the squared residual $|\hat A \ket{x} - \ket{b}|^2$ is smaller than $10^{-20}$~Ha$^2$ or~Ha$^4$, respectively.
Equation~(\ref{pe}) is calculated using 4 iterations of the Lanczos scheme described in Sec.~\ref{recyclage}.
Equations~(\ref{kge}) and~(\ref{khe}) are solved with the shift-Lanczos method~\cite{CB-13,Umari10}, where 8 and 16 iterations are used respectively. 
The parameter $\alpha$ of the Lorentzian model to the dielectric matrix (Eq.~(\ref{he})) is kept fixed at 1.0~Ha.
The cutoff energy is converged separately for the calculation of $\Sigma^{\rm x}_e$ and $\Sigma^{\rm c}_e(\delta)$ until the former is converged to 10~meV and the latter to 50~meV. 
The number of Lanczos vectors $N_L$ describing the exact part of the dielectric matrix $\hat \epsilon - \hat {\tilde \epsilon}$ (as described in Sec.~\ref{model}) and the number of Lanczos vectors $N_{\tilde L}$ describing the approximate dielectric matrix $\hat {\tilde \epsilon} - 1$ are also selected so that $\Sigma^{\rm c}_e(\delta)$ is converged to 50~meV, unless specified otherwise. 
When no value is given for $N_{\tilde L}$, the exact dielectric matrix $\hat \epsilon - 1$ is constructed using $N_L$ Lanczos vectors and the dielectric model is not used. 
The converged values for the box size, the cutoff energy and the number of Lanczos vectors ($N_L$ and $N_{\tilde L}$) are given for each molecule in Table~\ref{conv_par}.
\begin{table}
\caption[Converged values for molecule dependent parameters]{
Converged values for molecule dependent parameters. 
$\Sigma^{\rm x}_e$ cutoff and $\Sigma^{\rm c}_e$ cutoff are the cutoff energies used for $\Sigma^{\rm x}_e$ and $\Sigma^{\rm c}_e(\delta)$, respectively.
All quantities are given in atomic units.
}
\label{conv_par}
\begin{ruledtabular}
\begin{tabular}{lccccc}
molecule     & box size & $\Sigma^{\rm x}_e$ cutoff & $\Sigma^{\rm c}_e$ cutoff & $N_L$ & $N_{\tilde L}$ \\
\hline
silane       & 50       & 20                 & 20                 & 512   & -              \\
thiophene    & 26       & 40                 & 40                 & 512   & 2048           \\
benzene      & 28       & 30                 & 20                 & 240   & 1920           \\
naphthalene  & 30       & 30                 & 20                 & 384   & 3072           \\
anthracene   & 40       & 30                 & 20                 & 528   & 3168           \\
tetracene    & 51       & 30                 & 20                 & 504   & 4032           \\
C$_{60}$     & 40       & 40                 & 20                 & 1024  & 4096           \\
\end{tabular}
\end{ruledtabular}
\end{table}

\subsection{Silane}

\begin{table}
\caption[Comparison of the HOMO and LUMO of silane with the literature]{
Comparison between the present implementation and previously published $G_0W_0$ results for the highest occupied molecular orbital (HOMO) and the lowest unoccupied molecular orbital (LUMO) energies of the \textbf{silane} molecule. 
The underlying DFT energies are given to assess the agreement of the starting point of the $G_0W_0$ calculations.
All results are in eV.
}
\label{silane}
\begin{ruledtabular}
\begin{tabular}{lcccc}
\multicolumn{1}{l}{}              &\multicolumn{2}{c}{HOMO}      &\multicolumn{2}{c}{LUMO} \\
\hline
                                  & LDA     & $G_0W_0$         & LDA      & $G_0W_0$   \\
Ref.~\onlinecite{Rohlfing:1998fb} & $-$8.4  & $-$12.7            & -        & -            \\
Ref.~\onlinecite{Grossman:2001kc} & $-$8.4  & $-$12.7            & $-$0.6   & 0.3          \\
Ref.~\onlinecite{Hahn:2005ez}     & $-$8.42 & $-$12.41           & $-$0.50  & 0.50         \\
Ref.~\onlinecite{Bruneval:2012ii} & -       & $-$12.43           & -        & -            \\
This work                         & $-$8.51 & $-$12.43           & $-$0.53  & 0.79         \\
                                  &\multicolumn{2}{c}{Exp.}      &\multicolumn{2}{c}{Exp.} \\
Ref.~\onlinecite{Roberge:1978kb}  & \multicolumn{2}{c}{$-$12.3}  &\multicolumn{2}{c}{-}    \\
Ref.~\onlinecite{Itoh:1986ej}     & \multicolumn{2}{c}{$-$12.36} &\multicolumn{2}{c}{-}   \\
                                  & PBE     & $G_0W_0$         & PBE      & $G_0W_0$   \\
Ref.~\onlinecite{Setten13,*Setten13e}        & $-$8.47 & $-$12.11           & -        & -            \\
Ref.~\onlinecite{Bruneval:2012ii} & -       & $-$12.40           & -        & -            \\
Ref.~\onlinecite{Ren12}           & -       & $-$12.29           & -        & -            \\
Ref.~\onlinecite{vanSetten14}\footnote{
Result obtained with the FHI-aims code with a 16-parameter Pad\'e analytic continuation for the self-energy.}
                                  & -       & $-$12.31           & -        & 2.51         \\
Ref.~\onlinecite{vanSetten14}\footnote{
Result obtained with the TURBOMOLE code with no resolution-of-identity approximation.}
                                  & -       & $-$12.31           & -        & 2.51         \\
This work                         & $-$8.51 & $-$12.33           & $-$0.51  & 0.77         \\
\end{tabular}
\end{ruledtabular}
\end{table}

Results for silane are given in Table~\ref{silane}.
Good agreement is observed among the DFT results.
Indeed, while our DFT results are converged to a few meV, the other DFT studies were converged to 0.1~eV, which is comparable to the agreement between the results.
Similarly, good agreement is observed between the $G_0W_0$ results for the HOMO, except for Ref.~\onlinecite{Rohlfing:1998fb,Grossman:2001kc}.
Moreover, the comparison with experimental HOMO energies is also good.
These results thus support the accuracy of our implementation.

Interestingly, $G_0W_0$ results for the LUMO do not show such an agreement. 
Given the scatter of available $G_0W_0$ results and the absence of experiments, it becomes hard to gather information on the validity of our implementation from silane LUMO energies. 

\subsection{Thiophene}
\begin{table}
\caption[Comparison of the HOMO of thiophene with the literature]{ 
Results (in eV) for the HOMO energy of the \textbf{thiophene} molecule. 
}
\label{thiophene}
\begin{ruledtabular}
\begin{tabular}{lcc}
\multicolumn{1}{l}{}              &\multicolumn{2}{c}{HOMO}      \\
\hline
                                  & LDA     & $G_0W_0$         \\
Ref.~\onlinecite{Blase11}         & $-$6.15 & $-$8.37            \\
Ref.~\onlinecite{BlaseNote}\footnote{
Ref.~\onlinecite{BlaseNote} performs all-electron calculations with the Dunning aug-cc-pVTZ and aug-cc-pVQZ basis, instead of the much reduced double-zeta plus polarization (DZP) basis of Ref.~\onlinecite{Blase11}, but with the same Gaussian-basis contour-deformation methodology, and finds a $G_0W_0$ ionization energy for thiophene of 8.55 eV and 8.69 eV, respectively, with a starting LDA value of 6.06 eV with both basis. 
They conclude that these results indicate a slow convergence with respect to the size of the Gaussian basis in the case of small molecules with unbound virtual states. 
}
                                  & $-$6.06 & $-$8.55            \\
                                  & $-$6.06 & $-$8.69            \\
This work                         & $-$6.04 & $-$8.93            \\
                                  &\multicolumn{2}{c}{Exp.}      \\
Ref.~\onlinecite{Curtiss97}       & \multicolumn{2}{c}{$-$8.85}  \\
Ref.~\onlinecite{NIST:webbook}    & \multicolumn{2}{c}{$-$8.86$\pm$0.02}\\
                                  & PBE     & $G_0W_0$         \\
Ref.~\onlinecite{Sharifzadeh:2012}& -       & $-$9.0             \\
Ref.~\onlinecite{Pham:2013gs}     & $-$5.70 & $-$8.49            \\
This work                         & $-$5.86 & $-$8.73            \\
\end{tabular}
\end{ruledtabular}
\end{table}

The results for thiophene are given in Table~\ref{thiophene}.
Again, the difference between DFT results is reasonable, and gives an estimate of the expected agreement between $G_0W_0$ results, given the differences in boundary conditions, geometry, pseudopotentials and convergence criteria between studies.
In that respect, the agreement between $G_0W_0$ results is good excepted for Ref.~\onlinecite{Blase11}.
However, Ref.~\onlinecite{BlaseNote} repeated the calculations of Ref.~\onlinecite{Blase11} with larger localized basis sets and obtained a $G_0W_0$ result in good agreement with ours along with a slow convergence rate, as explained in the footnote of Table~\ref{thiophene}.
Together, these observations resolve this discrepancy. 
Finally, the agreement with experiments is good.

\subsection{Benzene}

\begin{table}
\caption[Comparison of the HOMO of benzene with the literature]{ 
Results (in eV) for the HOMO energy of the \textbf{benzene} molecule.  
}
\label{benzene}
\begin{ruledtabular}
\begin{tabular}{lcc}
\multicolumn{1}{l}{}                &\multicolumn{2}{c}{HOMO}      \\
\hline
                                    & LDA     & $G_0W_0$         \\
Ref.~\onlinecite{Tiago06}           & -       & $-$9.88            \\
Ref.~\onlinecite{productbasis}      & $-$6.67 & $-$8.78            \\
Ref.~\onlinecite{Umari10}           & -       & $-$9.40            \\
Ref.~\onlinecite{Ren12}             & -       & $-$9.05            \\
Ref.~\onlinecite{Samsonidze:2011bu} & $-$6.49 & $-$9.03            \\
Ref.~\onlinecite{Pham:2013gs}       & -       & $-$9.22            \\
This work                           & $-$6.50 & $-$9.23            \\
                                    &\multicolumn{2}{c}{Exp.}      \\
Ref.~\onlinecite{NIST:webbook}      & \multicolumn{2}{c}{$-$9.24378$\pm$0.00007} \\
                                    & PBE     & $G_0W_0$         \\
Ref.~\onlinecite{Sharifzadeh:2012}  & -       & $-$9.4             \\
Ref.~\onlinecite{Setten13,*Setten13e}          & $-$6.39 & $-$8.87            \\
Ref.~\onlinecite{vanSetten14}\footnote{
Result obtained with the FHI-aims code with a 16-parameter Pad\'e analytic continuation for the self-energy.}
                                    & -       & $-$8.99            \\
Ref.~\onlinecite{vanSetten14}\footnote{
Result obtained with the TURBOMOLE code with resolution-of-identity approximation.}
                                    & -       & $-$8.97            \\
Ref.~\onlinecite{Ren12}\footnote{
Result obtained with the numeric atom-centered orbitals set 'tier 4 + a5Z-\emph{d}'.}
                                    & -       & $-$9.00            \\
Ref.~\onlinecite{Pham:2013gs}       & $-$6.18 & $-$9.04            \\
This work                           & $-$6.31 & $-$9.03            \\
\end{tabular}
\end{ruledtabular}
\end{table}

Results for benzene are given in Table~\ref{benzene}.
Again, the agreement between DFT results is good.
We also find good agreement between $G_0W_0$ results (except for Ref.~\onlinecite{Tiago06}), which supports the accuracy of our implementation.
Moreover, our agreement with the experiment is also good.

\subsection{Naphthalene}
\begin{table}
\caption[Comparison of the HOMO of naphthalene with the literature]{ 
Results (in eV) for the HOMO energy of the \textbf{naphthalene} molecule. 
}
\label{naphthalene}
\begin{ruledtabular}
\begin{tabular}{lcc}
\multicolumn{1}{l}{}                &\multicolumn{2}{c}{HOMO}                \\ 
\hline
                                    & LDA     & $G_0W_0$                   \\ 
Ref.~\onlinecite{Tiago06}           & -       & $-$8.69                      \\ 
Ref.~\onlinecite{productbasis}      & -       & $-$7.67                      \\ 
This work                           & $-$5.67 & $-$8.05                      \\ 
                                    &\multicolumn{2}{c}{Exp.}                \\ 
Ref.~\onlinecite{NIST:webbook}      & \multicolumn{2}{c}{$-$8.144$\pm$0.001} \\ 
                                    & PBE     & $G_0W_0$                   \\ 
Ref.~\onlinecite{Setten13,*Setten13e}          & $-$5.50 & $-$7.73                      \\ 
This work                           & $-$5.48 & $-$7.84                      \\ 
\end{tabular}
\end{ruledtabular}
\end{table}

Results for naphthalene are given in Table~\ref{naphthalene}.
In this case, the agreement with the DFT result of Ref.~\onlinecite{Setten13,*Setten13e} is excellent. 
For the $G_0W_0$ results, the agreement with Ref.~\onlinecite{Setten13,*Setten13e} is also very good.
The other $G_0W_0$ results show reasonable but somewhat lesser agreement.
However, our results are in good agreement with the experiment.
This, together with the agreement with Ref.~\onlinecite{Setten13,*Setten13e}, supports the accuracy of our implementation. 

\subsection{Anthracene}

\begin{table}
\caption[Comparison of the HOMO and LUMO of anthracene with the literature]{
Results (in eV) for the HOMO and LUMO energies of the \textbf{anthracene} molecule.
}
\label{anthracene}
\begin{ruledtabular}
\begin{tabular}{lcccc}
\multicolumn{1}{l}{}                &\multicolumn{2}{c}{HOMO}      &\multicolumn{2}{c}{LUMO}  \\
\hline
                                    & LDA     & $G_0W_0$         & LDA      & $G_0W_0$    \\
Ref.~\onlinecite{Blase11}           & $-$5.47 & $-$6.89            & $-$3.22  & $-$0.74       \\
Ref.~\onlinecite{productbasis}      & -       & $-$6.89            & -        & $-$0.77       \\
Ref.~\onlinecite{Pham:2013gs}       & $-$5.18 & $-$7.25            & $-$2.81  & $-$1.05       \\
This work                           & $-$5.18 & $-$7.31            & $-$2.89  & $-$1.24       \\
                                    &\multicolumn{2}{c}{Exp.}      &\multicolumn{2}{c}{Exp.} \\
Ref.~\onlinecite{NIST:webbook}      & \multicolumn{2}{c}{$-$7.439$\pm$0.006} & \multicolumn{2}{c}{$-$0.66 to $-$0.42} \\
                                    & PBE     & $G_0W_0$         & PBE      & $G_0W_0$    \\
Ref.~\onlinecite{Setten13,*Setten13e}          & $-$4.96 & $-$7.12            & $-$2.68  & $-$1.17       \\ 
This work                           & $-$4.98 & $-$7.09            & $-$2.67  & $-$1.01       \\
\end{tabular}
\end{ruledtabular}
\end{table}
Results for anthracene are given in Table~\ref{anthracene}.
Both at the DFT and the $G_0W_0$ level, we observe good agreements with previously published results for the HOMO energies. 
Interestingly, we observe better agreement with Ref.~\onlinecite{Pham:2013gs} and Ref.~\onlinecite{Setten13,*Setten13e} than with Ref.~\onlinecite{Blase11}, similarly to the case of thiophene, and Ref.~\onlinecite{productbasis}, similarly to the case of benzene. 
We also observe good agreement between our HOMO $G_0W_0$ results and the experiment.

For LUMO DFT results, we also observe a good agreement with previously published results.
Still, the agreement between the LUMO $G_0W_0$ results is slightly diminished with respect to the HOMO, but this is due to the larger scatter of published data. 
Finally, the agreement with experiments is not very good for LUMO energies.
This is relatively unsurprising, since the accuracy of $G_0W_0$ results depends on the quality of the DFT starting point and LDA/PBE can produce particularly poor results for orbitals close to the vacuum level.  

\subsection{Tetracene}

\begin{table}
\caption[Comparison of the HOMO and LUMO of tetracene with the literature]{
Results (in eV) for the HOMO and LUMO energies of the \textbf{tetracene} molecule.
}
\label{tetracene}
\begin{ruledtabular}
\begin{tabular}{lcccc}
\multicolumn{1}{l}{}                &\multicolumn{2}{c}{HOMO}      &\multicolumn{2}{c}{LUMO}  \\
\hline
                                    & LDA     & $G_0W_0$         & LDA      & $G_0W_0$    \\
Ref.~\onlinecite{Blase11}           & $-$5.15 & $-$6.37            & $-$3.58  & $-$1.34       \\
Ref.~\onlinecite{Pham:2013gs}       & $-$4.85 & $-$7.04            & $-$3.19  & $-$1.41       \\
This work                           & $-$4.86 & $-$6.79            & $-$3.26  & $-$1.80       \\
                                    &\multicolumn{2}{c}{Exp.}      &\multicolumn{2}{c}{Exp.} \\
Ref.~\onlinecite{NIST:webbook}      & \multicolumn{2}{c}{$-$6.97$\pm$0.05} & \multicolumn{2}{c}{$-$1.06 to $-$0.88} \\
                                    & PBE     & $G_0W_0$         & PBE      & $G_0W_0$    \\
Ref.~\onlinecite{Setten13,*Setten13e}          & $-$4.65 & $-$6.70            & $-$3.05  & $-$1.84       \\ 
This work                           & $-$4.66 & $-$6.57            & $-$3.04  & $-$1.56       \\
\end{tabular}
\end{ruledtabular}
\end{table}
Results for tetracene are given in Table~\ref{tetracene}.
Again, we observe our results to be in good agreement with previously published results at the DFT level.
At the $G_0W_0$ level, for HOMO energies, the agreement with Ref.~\onlinecite{Setten13,*Setten13e} is good, as in the cases of naphthalene and anthracene. 
Also, the agreement with Ref.~\onlinecite{Blase11} is similar to the one observed in the case of thiophene, which is coherent with the uniform basis size they used to simulate all their molecules (see footnote of Table~\ref{thiophene}).
However, the agreement with Ref.~\onlinecite{Pham:2013gs} is somewhat lesser than expected.
Given that our agreement with Ref.~\onlinecite{Pham:2013gs} is good for all other molecules investigated in this study, it is plausible that the difference stems from the choice of simulation parameters and not the implementation. 
Moreover, we observe reasonable agreement between our HOMO $G_0W_0$ results and the experiment. 

Available $G_0W_0$ data is more scattered for the LUMO than the HOMO, but the agreement remain otherwise similar, i.e. we agree best with Ref.~\onlinecite{Setten13,*Setten13e}, then Ref.~\onlinecite{Pham:2013gs} and finally Ref.~\onlinecite{Blase11}.
Also, due to the poor description of the weakly bound LUMO orbital by LDA/PBE, the LUMO $G_0W_0$ results are not in very good agreement with experiment.

\subsection{C$_{60}$}

\begin{table}
\caption[Comparison of the HOMO and LUMO of C$_{60}$ with the literature]{
Results (in eV) for the HOMO and LUMO energies of the \textbf{C$_{60}$} molecule.
}
\label{C60}
\begin{ruledtabular}
\begin{tabular}{lcccc}
\multicolumn{1}{l}{}                   &\multicolumn{2}{c}{HOMO}      &\multicolumn{2}{c}{LUMO}  \\
\hline
                                       & PBE     & $G_0W_0$         & PBE      & $G_0W_0$    \\
Ref.~\onlinecite{Samsonidze:2011bu}    & $-$5.84 & $-$7.21            & $-$4.19  & $-$2.62       \\
Ref.~\onlinecite{Pham:2013gs}          & $-$5.81 & $-$7.31            & $-$4.13  & $-$2.74       \\
This work                              & $-$5.84 & $-$7.41            & $-$4.15  & $-$2.92       \\
                                       &\multicolumn{2}{c}{Exp.}      &\multicolumn{2}{c}{Exp.} \\
Ref.~\onlinecite{Lichtenberger:1991bw} & \multicolumn{2}{c}{$-$7.6$\pm$0.2} & \multicolumn{2}{c}{-} \\
Ref.~\onlinecite{Huang:2014ha}         & \multicolumn{2}{c}{-} & \multicolumn{2}{c}{$-$2.684$\pm$0.007} \\
\end{tabular}
\end{ruledtabular}
\end{table}
Results for C$_{60}$ are given in Table~\ref{C60}.
All DFT and $G_0W_0$ results show good agreement. 
We also observe a reasonable agreement with experiment.

\subsection{Analysis}

Globally, our $G_0W_0$ results for the HOMO of all molecules as well as the LUMO of anthracene, tetracene, and C$_{60}$ support the validity and accuracy of the present implementation.
The $G_0W_0$ results for the LUMO of silane are, however, harder to analyze, due to the large scatter in available data. 
Still, this latter point has no consequence in the assessment of our implementation and, therefore, this study clearly support the implementation's accuracy.

\begin{table}
\caption[Contributions to $G_0W_0$ eigenenergies]{
Contributions to $G_0W_0$ eigenenergies $\varepsilon^{G_0W_0}_e$. 
The DFT eigenenergy obtained using BigDFT is noted $\varepsilon^{\rm DFT}_e$, while $V^{\rm xc}_e$, $\Sigma^{\rm x}_e$ and $\Sigma^{\rm c}_e$ stand for the expectation value of the corresponding operator in the state considered $\ket{e}$.
$\Sigma^{\rm c}_e$ is evaluated at the energy solving Eq.~\eqref{exc} when the ABINIT DFT eigenenergy is used.
}
\label{contributions}
\begin{ruledtabular}
\begin{tabular}{lccccccc}
molecule    & orbital & funct.     & $\varepsilon^{\rm DFT}_e$ & $V^{\rm xc}_e$ & $\Sigma^{\rm x}_e$ & $\Sigma^{\rm c}_e$ & $\varepsilon^{G_0W_0}_e$ \\
\hline
sil.        & HOMO    & LDA        & $-$8.51                   & $-$10.98       & $-$15.61           &    0.71            & $-$12.43                 \\
            &         & PBE        & $-$8.51                   & $-$11.29       & $-$15.82           &    0.71            & $-$12.33                 \\
            & LUMO    & LDA        & $-$0.53                   & $-$2.90        & $-$0.92            & $-$0.67            &    0.79                  \\
            &         & PBE        & $-$0.51                   & $-$2.68        & $-$0.81            & $-$0.59            &    0.77                  \\
thio.       & HOMO    & LDA        & $-$6.04                   & $-$13.03       & $-$15.75           & $-$0.17            & $-$8.93                  \\
            &         & PBE        & $-$5.86                   & $-$13.09       & $-$15.79           & $-$0.16            & $-$8.73                  \\
benz.       & HOMO    & LDA        & $-$6.50                   &  $-$12.97      &    $-$15.54        & $-$0.16            & $-$9.23                  \\
            &         & PBE        & $-$6.31                   &  $-$13.01      &    $-$15.58        & $-$0.15            & $-$9.03                  \\
napht.      & HOMO    & LDA        & $-$5.67                   &  $-$13.13      &   $-$15.14         & $-$0.36            & $-$8.05                  \\
            &         & PBE        & $-$5.48                   &  $-$13.17      &   $-$15.18         & $-$0.35            & $-$7.84                  \\
anthr.      & HOMO    & LDA        & $-$5.18                   &  $-$13.23      &   $-$14.84         & $-$0.52            & $-$7.31                  \\
            &         & PBE        & $-$4.98                   &  $-$13.26      &   $-$14.86         & $-$0.51            & $-$7.09                  \\
            & LUMO    & LDA        & $-$2.89                   &  $-$13.03      &   $-$8.80          & $-$2.59            & $-$1.24                  \\
            &         & PBE        & $-$2.67                   &  $-$13.05      &   $-$8.81          & $-$2.58            & $-$1.01                  \\
tetr.       & HOMO    & LDA        & $-$4.86                   &  $-$13.31      &  $-$14.62          & $-$0.61            & $-$6.79                  \\
            &         & PBE        & $-$4.66                   &  $-$13.36      &  $-$14.66          & $-$0.60            & $-$6.57                  \\
            & LUMO    & LDA        & $-$3.26                   &  $-$13.08      &  $-$9.12           & $-$2.49            & $-$1.80                  \\
            &         & PBE        & $-$3.04                   &  $-$13.11      &  $-$9.15           & $-$2.48            & $-$1.56                  \\
C$_{60}$    & HOMO    & PBE        & $-$5.84                   & $-$13.85       & $-$15.23           & $-$0.18            & $-$7.41                  \\
            & LUMO    & PBE        & $-$4.15                   & $-$13.56       & $-$9.81            & $-$2.52            & $-$2.92                  \\
\end{tabular}
\end{ruledtabular}
\end{table}
To enable further analysis, the decomposition of $G_0W_0$ eigenenergies into their different contributions is given in Table~\ref{contributions}.
It is interesting to note that changing the functional has little effect on the value of $\Sigma^{\rm c}_e$ for all molecules studied. 
Thus, the choice of functional affects $\varepsilon^{G_0W_0}_e$ through the computationally simpler contributions $\varepsilon^{\rm DFT}_e$, $V^{\rm xc}_e$ and $\Sigma^{\rm x}_e$.

Also, for molecules containing only the first row elements, changing LDA for PBE had the same effect on all the DFT and $G_0W_0$ eigenenergies: it raised them by about 0.2~eV.
Therefore, the $G_0W_0$ correction $\Delta \varepsilon_e = \bra{e} \hat \Sigma(\varepsilon_e + \Delta \varepsilon_e) - \hat V^{\rm xc} \ket{e}$ is quite insensitive to the choice of functional for these molecules.
However, since their HOMOs are qualitatively similar (in that they are formed of delocalized $\pi$ orbitals), the full range of molecules to which this trend applies is difficult to guess.

\section{Performance}
\label{performance}
Since the scaling of our implementation is the same as the traditional $G_0W_0$ implementation~\cite{Hybertsen86} ($\propto N^4$ in both cases), it becomes interesting to assess the prefactor by direct comparison of the computation times for both methods.
To this end, we calculate the expectation value of the correlation part of the self-energy for the HOMO orbital of the silane molecule $\Sigma^{\rm c}_e(0)$ (see Eq.~(\ref{e_c1})) with both our implementation and the traditional $G_0W_0$ implementation found in the ABINIT project~\cite{Gonze:2009aa}.
To allow a direct comparison of the computation times, we keep all parameters to their converged value, except for the unit cell size, the cutoff energy, the number of Lanczos vectors $N_L$ and the number of plane-waves used to describe the dielectric matrix $N_{PW\epsilon}$.
The unit cell size and the cutoff energy are common to both calculations and are kept to an underconverged value of 18~bohr and 4~Ha respectively, to allow the conventional $G_0W_0$ calculations to fit on the memory available on our computers. 
Since the number of Lanczos vectors $N_L$ and the number of plane-waves used to describe the dielectric matrix $N_{PW\epsilon}$ are analogous parameters, i.e. they control the precision of the dielectric operator in the present and the conventional $G_0W_0$ implementation, respectively, they are chosen as convergence parameters. 
No dielectric model was used in the present implementation, so that the dimension of the dielectric matrix remains controlled by one unique parameter. 
The ratio of the prefactors is then obtained from the ratio of the computation times required for both implementations to reach a given level of convergence.
Here, a level of convergence of $\pm50$~meV is chosen, since it is the typical expected accuracy of $G_0W_0$ calculations. 

\begin{figure}
\includegraphics[width=1.0 \linewidth]{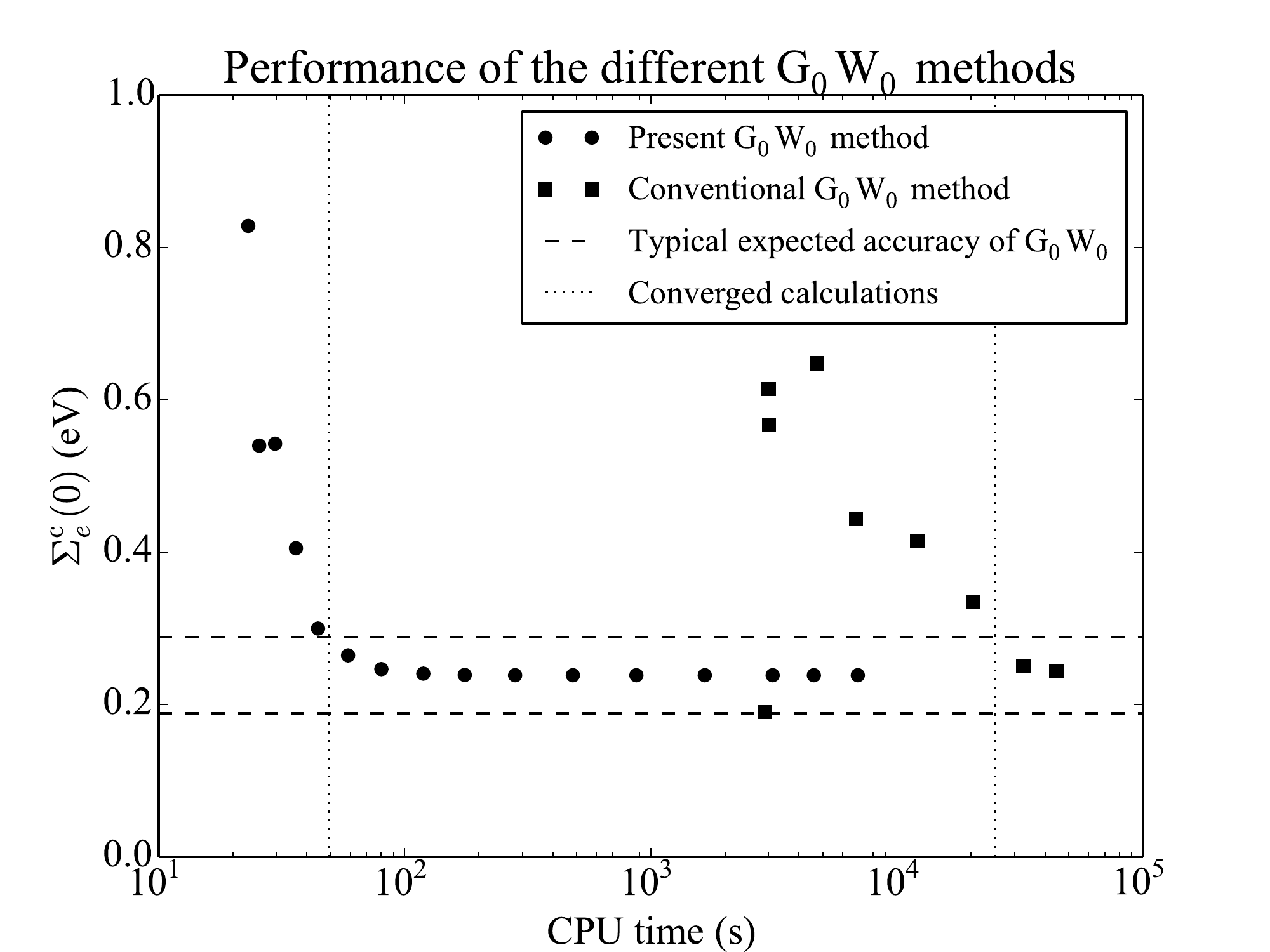}  
\caption[Convergence study on the expectation value of the correlation part of the self-energy for the HOMO orbital of the silane molecule $\Sigma^{\rm c}_e(0)$  using the present and the conventional $G_0W_0$ implementation of the ABINIT project]{ \label{cput} Convergence study on the expectation value of the correlation part of the self-energy for the HOMO orbital of the silane molecule $\Sigma^{\rm c}_e(0)$  using the present and the conventional $G_0W_0$ implementation of the ABINIT project. 
The convergence studies are carried out with respect to the size of the dielectric matrix, which is controlled by the number of Lanczos vectors $N_L$ in the first implementation and the number of plane-waves used to describe the dielectric matrix $N_{PW\epsilon}$ in the second.
The horizontal lines show the energy zone considered to be converged and the vertical lines show the approximate CPU time required to reach this level of convergence, based on a linear interpolation between the first data point to be converged and the preceding one.
}
\end{figure}
The results obtained for the convergence studies are shown in Fig.~\ref{cput}.
This analysis shows that our implementation is about 500 times faster than the conventional one for this calculation size.

Although the present implementation does not yet support periodic systems, it is still possible to extend further the performance analysis by doing (unphysical) simulations of crystals where only one $k$ point is used to sample the Brillouin zone ($\Gamma$). 
We carry out such simulations for silicon, diamond and graphite.
All parameters are kept to a common, converged value, except for the number of Lanczos vectors $N_L$ in the present implementation as well as the number of conduction states $N_c$ and the number of plane-waves used to describe the dielectric matrix $N_{PW\epsilon}$ in the conventional implementation. 
For the present implementation, we choose $N_L$ so that the result is converged to 50~meV, while for the conventional implementation, we choose $N_c$ and $N_{PW\epsilon}$ so that they each contribute 25~meV to the final error, the latter choice being generally the most computationally efficient way to reach a final error of 50~meV. 
These tests show our implementation to be five times, six times, and 20 times faster than the conventional implementation for silicon, diamond, and graphite, respectively.
Thus, the speedup of the present implementation increases as the number of plane-waves per valence electron required to accurately describe the dielectric matrix increases, silicon likely being one of the lowest possible speedup. 

Furthermore, this speedup is for an unphysical simulation where only one $k$ point is used to sample the Brillouin zone.
Indeed, the generalization of our implementation to periodic systems should profit from re-use of computationally expensive information (such as the Lanczos basis $\{\ket{l}\}$ or the bases $\{\ket{\gamma_{i,v}}\}$ used to solve Eq.~\eqref{P2+} at finite frequency) from one $k$ point to another, unlike the conventional implementation, which requires equally large computational effort for all $k$ points. 
The speedup obtained here for extended systems should therefore be substantially lower than those to be reached with the extension of our implementation to periodic systems, by a factor of up to the number of $k$ points used to sample the Brillouin zone.

\section{Conclusion}
The $G_0W_0$ implementation presented here successfully circumvents the two bottlenecks present in conventional plane-wave implementations.
The conversion of the summations over conduction states into Sternheimer equations effectively eliminates the first bottleneck.
The second one is solved by expressing the dielectric matrix in a Lanczos basis.
This effectively reduces its size by orders of magnitude, to the level of spectral decomposition methods~\cite{Pham:2013gs}, while being computationally an order of magnitude cheaper than the latter.
Also, we developed a model dielectric operator, which further reduces the size of the dielectric matrix without accuracy loss.
Furthermore, we explored two ways to alleviate the computational cost of the integration over frequencies without resorting to approximations such as the plasmon pole model.
First, a scalar Lorentzian model for the frequency dependence of the dielectric matrix on the imaginary axis is used to reduce the frequency sampling required to evaluate the integral.
This particular model has the advantages of having a simple physical interpretation, presenting the right high-frequency behavior and being compatible with the conversion of summations over conduction states into linear equations.
Also, we use a scheme that provides the dielectric matrix at any imaginary frequency for a negligible computational cost, based on the recycling of the information computed in the construction of the static dielectric matrix.
The latter two concepts reduce the computational cost of the integration over frequencies by an order of magnitude.
This, combined with the elimination of the bottlenecks mentioned previously, effectively reduces the computation time required to achieve convergence by orders of magnitude with respect to a conventional plane-wave $G_0W_0$ implementations. 
Small tests for the silane molecule revealed a 500-fold speedup. 

This reduction in computational cost is achieved while preserving the high numerical precision provided by plane-wave basis sets.
Indeed, this implementation uses the contour deformation technique at almost no additional cost with respect to plasmon pole approximations, thanks to the use of the Lorentzian model and the recycling scheme.
Also, the conversion of the summations over conduction states into linear equation problems eliminates the need to converge the results with respect to the number of states included in these notoriously slow converging summations, thus eliminating a numerical source of uncertainty.
Moreover, the natural ability of the Lanczos method to first explore the biggest contributions of the dielectric matrix to the $G_0W_0$ results smooths the convergence behavior with respect to the matrix size.
Finally, the use of the SQMR algorithm for the iterative solution of linear equation problems allows the $G_0W_0$ corrections to be computed directly at the desired real frequency, thus avoiding the need for analytic continuations and related stability considerations. 

For all molecules considered in this work, the computed quasiparticle energies show good agreement with previously published $G_0W_0$ results (except for the LUMO of silane, where the scatter of available data makes the analysis difficult), thus validating the accuracy of our implementation. 
Moreover, our agreements with experimental ionization energies is similar or better than those of previously published $G_0W_0$ studies, suggesting that this implementation is effective in preserving the full precision of the $G_0W_0$ method.
Also, though the results presented here are for molecules, the extension of our method to crystals (using $k$ points grids) is straightforward and under way. 

\section{Acknowledgments}
We would like to thank Xavier Blase and Steven Louie for helpful discussions.
We would also like to thank Yann Pouillon and Jean-Michel Beuken for their valuable technical support and help with the test and build system of ABINIT.
Moreover, we acknowledge the Natural Sciences and Engineering Research Council of Canada, the Fonds de recherche du Qu\'ebec - Nature et technologies, the Regroupement qu\'eb\'ecois sur les mat\'eriaux de pointe, and the Photovoltaic Innovation Network for funding.
Finally, we acknowledge Calcul Qu\'ebec and Compute Canada for computational resources.

\bibliography{references_spec,references_v3}

\end{document}